# GeoSQL-Eval: First Evaluation of LLMs on PostGIS-Based NL2GeoSQL Queries

Shuyang Hou[a], Haoyue Jiao[b]*, Ziqi Liu[a], Lutong Xie[a], Guanyu Chen[b], Shaowen Wu[a], Xuefeng Guan [a], Huayi Wu[a]

a. State Key Laboratory of Information Engineering in Surveying, Mapping, and Remote Sensing, Wuhan University, Wuhan, China

*b. School of Resource and Environmental Sciences, Wuhan University, Wuhan, China

*Corresponding author: Haoyue Jiao, email: haoyuejiao@whu.edu.cn

**Abstract**

Large language models (LLMs) have shown strong performance in natural language to SQL (NL2SQL) tasks within general databases. However, extending to GeoSQL introduces additional complexity from spatial data types, function invocation, and coordinate systems, which greatly increases generation and execution difficulty. Existing benchmarks mainly target general SQL, and a systematic evaluation framework for GeoSQL is still lacking. To fill this gap, we present **GeoSQL-Eval**, the first end-to-end automated evaluation framework for PostGIS query generation, together with **GeoSQL-Bench**, a benchmark for assessing LLM performance in NL2GeoSQL tasks. GeoSQL-Bench defines three task categories—conceptual understanding, syntax-level SQL generation, and schema retrieval—comprising 14,178 instances, 340 PostGIS functions, and 82 thematic databases. GeoSQL-Eval is grounded in Webb's Depth of Knowledge (DOK) model, covering four cognitive dimensions, five capability levels, and twenty task types to establish a comprehensive process from knowledge acquisition and syntax generation to semantic alignment, execution accuracy, and robustness. We evaluate 24 representative models across six categories and apply the entropy weight method with statistical analyses to uncover performance differences, common error patterns, and resource usage. Finally, we release a public **GeoSQL-Eval leaderboard platform** for continuous testing and global comparison. This work extends the NL2GeoSQL paradigm and provides a standardized, interpretable, and extensible framework for evaluating LLMs in spatial database contexts, offering valuable references for geospatial information science and related applications.

## 1. Introduction

Relational databases have long served as the dominant solution for data management and querying, owing to their structured data modeling capabilities, transaction consistency guarantees, and strong scalability[1, 2]. Structured Query Language (SQL), as the standardized interface, is widely employed for general-purpose data operations within relational database systems[3]. In recent years, the rapid development of heterogeneous data acquisition technologies—including sensor networks, mobile devices, social media platforms, and remote sensing satellites—has produced increasingly large volumes of data enriched with spatial and temporal attributes[4-6]. Such spatiotemporal data not only enhance semantic dimensions but also introduce diverse geometric data types (e.g., points, line strings, polygons) and computational requirements for spatial analysis[7], such as topological relationship evaluation, buffer generation, spatial joins, distance calculations, and geometric transformations. Traditional relational databases, however, exhibit limitations in geometric data support, spatial indexing structures, and spatial function expressiveness, making them inadequate for real-world applications in urban planning, environmental monitoring, and traffic management[8-10]. To address these challenges, mainstream database systems have released spatial extensions, including MySQL Spatial Extensions, Oracle Spatial and Graph, SQLite with SpatiaLite, and PostgreSQL with its spatial module PostGIS[11, 12]. PostGIS has achieved broad adoption in the open-source spatial database domain, largely due to its comprehensive spatial function library, compliance with OGC standards, open-source architecture, and flexible on-premise deployment[13]. According to the 2025 DB-Engines ranking of geospatial databases, PostgreSQL holds a 48.7% market share in the open-source spatial database sector. Its spatial extension, PostGIS, has been widely deployed in smart city development, intelligent transportation, and geospatial information services, establishing itself as one of the most representative system platforms for GeoSQL query tasks[14, 15].

Spatial database extensions such as PostGIS primarily comply with standards established by the Open Geospatial Consortium (OGC), including the Simple Feature Access (SFA) model and the Geography Markup Language (GML), as well as international geoinformation standards such as ISO 19125[16, 17]. Building on the general SQL syntax and data type system, these extensions introduce a variety of spatial data types (e.g., GEOMETRY, GEOGRAPHY), spatial functions (e.g., ST_Contains, ST_DWithin), spatial indexing mechanisms (e.g., GiST), and spatial-specific query operators (e.g., &&, ~=). In addition, PostGIS supports multiple formats for representing and storing spatial data, including text-based representations (WKT) and binary encodings (WKB), thereby addressing diverse requirements for data transmission and processing[18, 19]. Compared with conventional SQL queries, GeoSQL not only incorporates spatial data types and function calls at the syntactic level but also requires users to understand fundamental spatial concepts such as spatial reference systems (SRID), spatial precision, topological relationships, and the logic of spatial operations and analysis. Consequently, although PostGIS offers highly flexible and comprehensive spatial analytical capabilities, the usability threshold varies considerably depending on the user's background. The implementation of Text2GeoSQL is essentially a cross-level transformation from semantics to execution. It entails establishing mappings between a user's natural language queries and the database's structured schema, while incorporating geospatial reasoning knowledge to refine and constrain semantics. This process produces GeoSQL queries that are both syntactically and semantically valid, returning structured outputs and spatial visualizations upon execution, thereby forming an integrated workflow that encompasses semantic parsing, query generation, and result presentation[20, 21], as illustrated in **Figure 1a**.

Users with different academic and professional backgrounds encounter distinct challenges when learning and applying PostGIS. For users from computer science backgrounds, SQL syntax is generally familiar, and learning the PostGIS-specific syntax poses relatively little difficulty. However, without an understanding of spatial data semantics, they may struggle to construct correct query logic, particularly when applying spatial analysis methods, interpreting spatial data characteristics, handling projections and coordinate transformations, managing spatial precision, and maintaining topological consistency. By contrast, users with backgrounds in geographic information systems (GIS) often have a clearer grasp of spatial concepts and analytical logic, but in the absence of SQL programming skills and database operation experience, they must additionally acquire knowledge of query languages, spatial data modeling, and function invocation methods. This results in a steep learning curve and reduced efficiency. For users engaged in interdisciplinary research who seek to leverage spatial databases, the challenges are even greater, as they often lack both programming expertise and geospatial reasoning skills, making it difficult to construct spatial queries and analytical workflows in PostGIS. The scarcity of relevant learning resources further exacerbates these barriers, leading to high costs in knowledge transfer and application. Even users proficient in both SQL and geospatial reasoning may encounter substantial time and resource consumption due to the logical complexity and human errors inherent in manual coding. As illustrated in **Figure 1b**, these challenges vary significantly across different user groups in the context of GeoSQL.

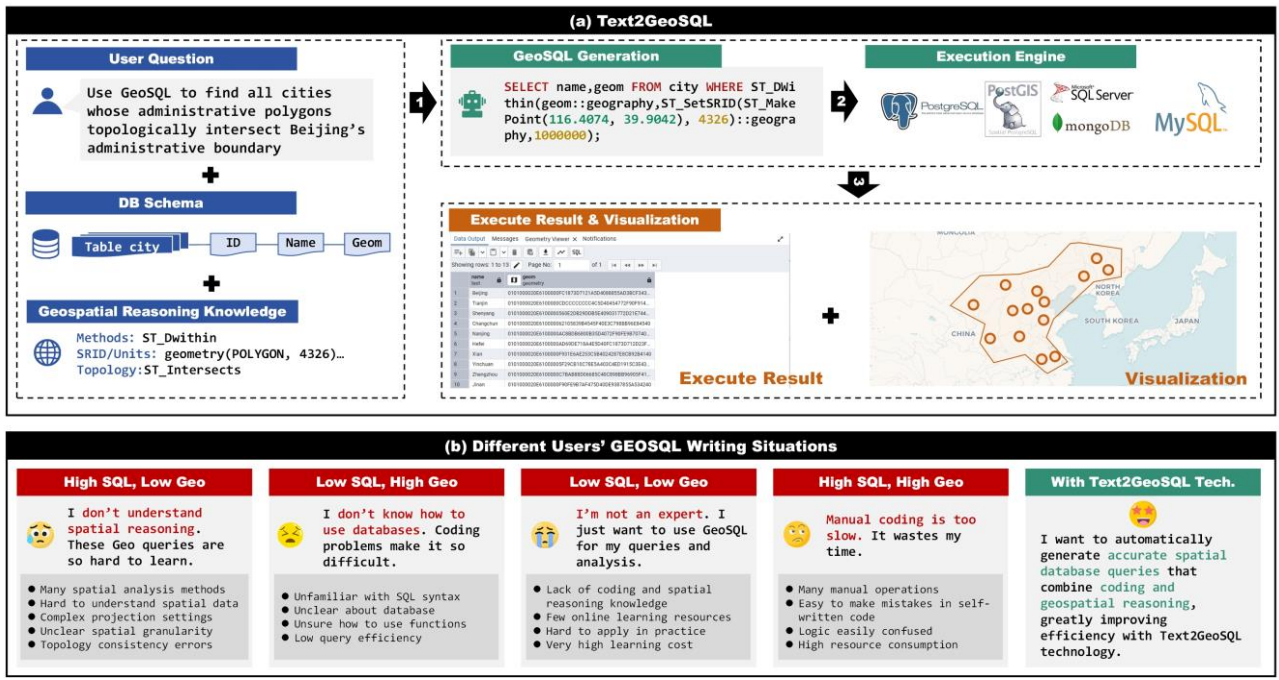

**Figure 1** Overview of Text2GeoSQL Workflow and User Scenarios.

In recent years, large language models (LLMs) have achieved continuous breakthroughs in natural language question answering, probabilistic language modeling, and instruction following, and they have been widely applied to the task of generating structured queries from natural language (Text-to-SQL or NL2SQL). Prior studies demonstrate that LLMs can extract query intent from natural language questions and produce SQL statements that are semantically coherent and structurally complete[22, 23]. However, when extended to GeoSQL systems such as PostGIS, the generation process faces substantially greater complexity. Although the output may appear semantically plausible and even pass syntactic validation by a parser, execution often fails due to latent errors or produces results inconsistent with expectations, as illustrated in **Figure 2**. One typical issue is Function Hallucination, where the model fabricates function names that do not exist. For instance, PostGIS does not include functions such as *ST_Makecircularstring* or *ST_BufferDistance*. While these names are semantically interpretable and conform to naming conventions, they are not valid functions, thereby rendering the query inoperable. The second common problem is Parameter Misuse, which encompasses errors in parameter order or type (e.g., applying *ST_DWithin(geog, dist)* to a GEOMETRY type, or misplacing the distance argument), missing parameters (e.g., omitting a displacement argument in *ST_Translate(geom, dx, dy)*), invalid parameter values (e.g., a negative or string-based buffer radius in *ST_Buffer*), and ambiguities caused by function overloading (e.g., invoking *ST_SetSRID('POINT(1 2)', 4326)* without explicitly casting *'POINT(1 2)'::geometry*, which prevents disambiguation among candidate functions). A third category involves Invalid Geometry Construction, where models generate geometries that violate WKT/WKB specifications. Examples include missing closure rings in TIN or POLYGON geometries, producing errors such as "geometry contains non-closed rings," or mixing 2D and 3D coordinates within a single object, resulting in "cannot mix dimensionality." The fourth issue is GeoSQL Syntax Misuse, such as incorrectly attempting to access fields from a set result with *ST_Dump(geom).geom*, which is not supported by SQL syntax. The fifth category relates to SRID and Projection Errors, including mismatched SRIDs between objects, references to nonexistent SRIDs, or the generation of invalid proj4 definitions, all of which cause failures in the PROJ library. Beyond these, certain functions in PostGIS, while syntactically valid, exhibit Environment-Dependent Behavior. Their outcomes may vary depending on specific configurations such as PostGIS version, support for three-dimensional (Z) coordinates, default SRID settings, or the strictness of OGC compliance. For example, ST_Union can yield different results when processing MULTIPOLYGON objects with Z values, leading to discrepancies in geometric precision and conformance. Most of these potential errors are difficult to detect through semantic parsers or static syntax checks. Without sufficient knowledge of PostGIS function interfaces and geospatial reasoning, users often struggle to promptly identify and correct such issues, thereby diminishing the usability and reliability of the generated queries. Ironically, the natural language interface—designed to lower entry barriers—may instead impose additional burdens in spatial query tasks. Moreover, due to data privacy requirements, some users prefer locally deployable open-source small models to ensure data security, whereas in resource-constrained environments or restricted deployment settings, users may opt for mainstream proprietary large models accessed via APIs to enable GeoSQL generation. Consequently, against the backdrop of expanding geospatial database applications, systematically evaluating the capabilities of current mainstream LLMs in generating PostGIS queries—clarifying their performance, error types, and applicability boundaries—holds significant value for users across domains such as computer science, geographic information science, urban studies, and transportation analysis.

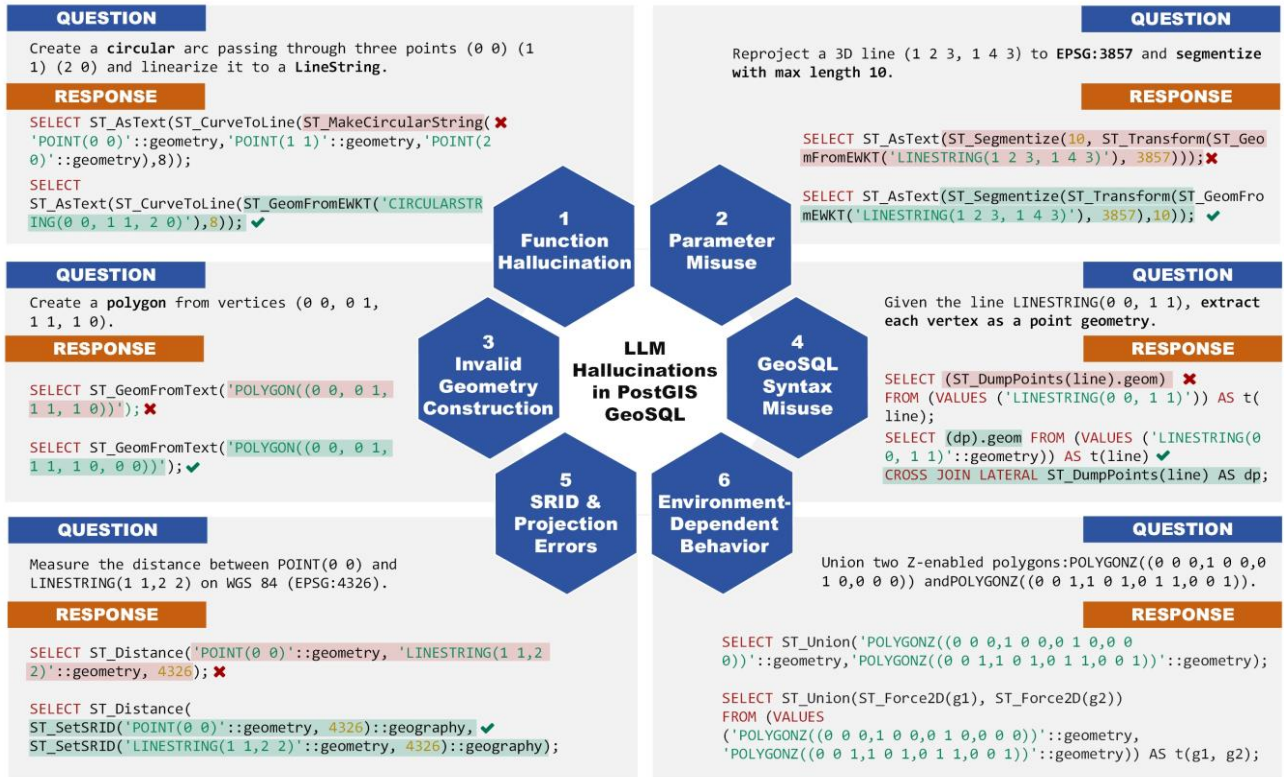

**Figure 2** Types of LLM Hallucinations in PostGIS GeoSQL

In the domain of general SQL, benchmark datasets such as Spider, WikiSQL, and BIRD have been widely adopted for standardized performance evaluation in NL2SQL tasks[24-26]. These datasets primarily focus on semantic mapping and syntactic generation within relational databases, without addressing the spatial semantics, spatial functions, or geometric data types that are unique to GeoSQL, nor the structural and syntactic features specific to PostGIS. In the field of geospatial computing, some studies have made preliminary attempts to evaluate the performance of LLMs in spatial code generation tasks, including GeoCode-Eval, GeoCode-Bench, the AutoGEEval series (AutoGEEval and AutoGEEval++), and GeoJSEval[27-30]. The evaluation tasks developed in these studies mainly target the generation of remote sensing image processing scripts or Google Earth Engine (GEE) code, covering image computation workflows and spatial process modeling, but they do not address SQL query construction in spatial databases. Other efforts, such as SPOK and GeoGPT, design evaluation tasks to test the spatial structural perception of LLMs, focusing on the interpretation and representation of intermediate structures such as WKT expressions, spatial annotations, geometric encodings, and spatial relationship vectors[31, 32]. Although these tasks involve spatial semantics, they do not incorporate the full process of GeoSQL query generation and lack comprehensive evaluation at the syntactic, structural, and execution levels. At present, only two studies have preliminarily explored LLM evaluation in geography-related SQL generation. The first is the GeoQueryJP dataset, which evaluates models' disambiguation ability for place names in the Japanese geographic context[33]. While geographic background information is included, the generated queries are still based on general SQL and do not cover GeoSQL-specific functions and data types. The second is the SpatialSQL benchmark constructed on

SpatiaLite, which includes four domains—administrative divisions (Ada), education (Edu), tourism (Tourism), and transportation (Traffic)—and consists of approximately 200 natural language queries supporting basic spatial distance calculations and entity filtering[21]. Although this benchmark introduces spatial query elements, its underlying database structure is relatively simplified, lacks PostGIS-specific functions and multidimensional structures, and limits evaluation metrics to correctness of results, without systematic assessment of syntactic validity, function invocation structure, schema linkage consistency, or robustness. In summary, there is currently no benchmark framework targeting PostGIS that can address practical requirements in structural generation, semantic alignment and invocation, and task complexity. Developing a GeoSQL benchmark that systematically covers cognitive understanding, syntactic generation, execution validation, semantic recognition and schema linkage, and robustness analysis is of clear research significance for evaluating the generative capacity and applicability boundaries of LLMs in spatial database contexts.

Building on the above background, this study adopts Norman L. Webb's Depth of Knowledge (DOK) [34] framework as its theoretical foundation and introduces **GeoSQL-Eval**, the first end-to-end automated evaluation framework designed to systematically assess the performance of LLMs in PostGIS query generation tasks. In parallel, we construct the first benchmark dataset, **GeoSQL-Bench**, which spans the entire generation pipeline from language understanding to query execution. The benchmark comprises four cognitive dimensions, five progressive capability levels, and twenty categories of evaluation tasks, enabling comprehensive and stratified measurement of the GeoSQL generation process. Specifically, the four cognitive dimensions align with the four levels of cognitive depth defined by the DOK model, while the five capability levels advance sequentially to establish a systematic evaluation framework ranging from knowledge acquisition to generalized reasoning. The overall evaluation framework is illustrated in **Figure 3**, and the end-to-end workflow of this study is depicted in **Figure 4**.

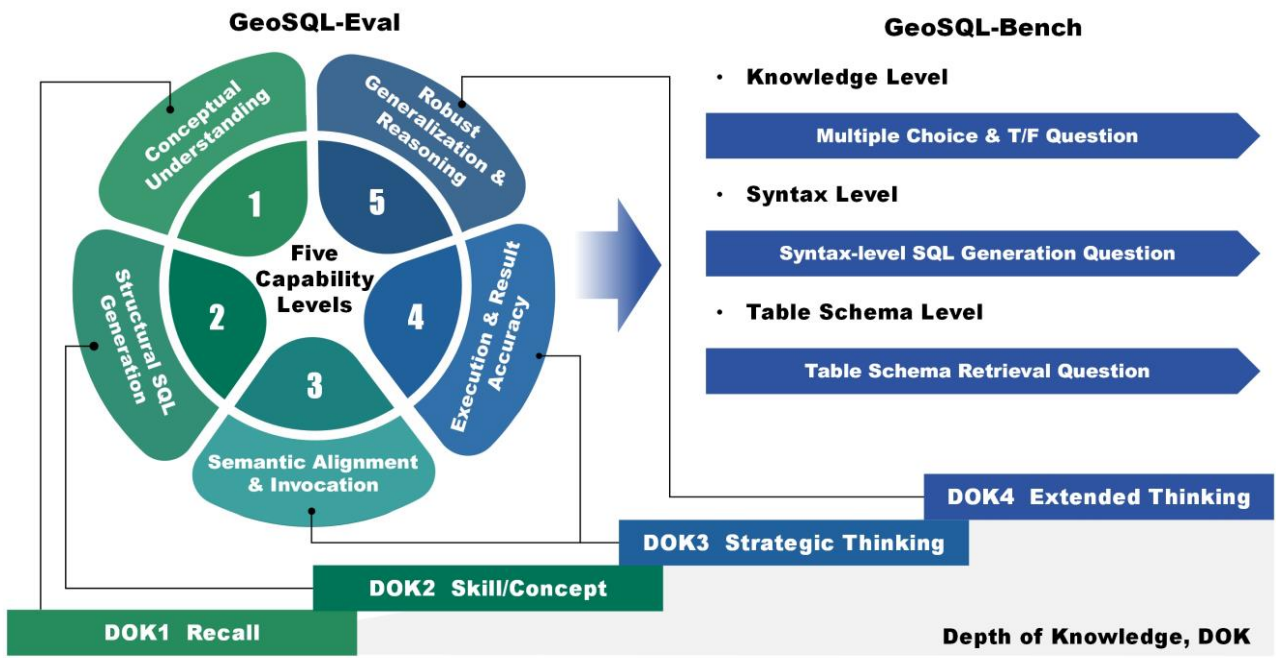

**Figure 3** Diagram of GeoSQL-eval evaluation framework.

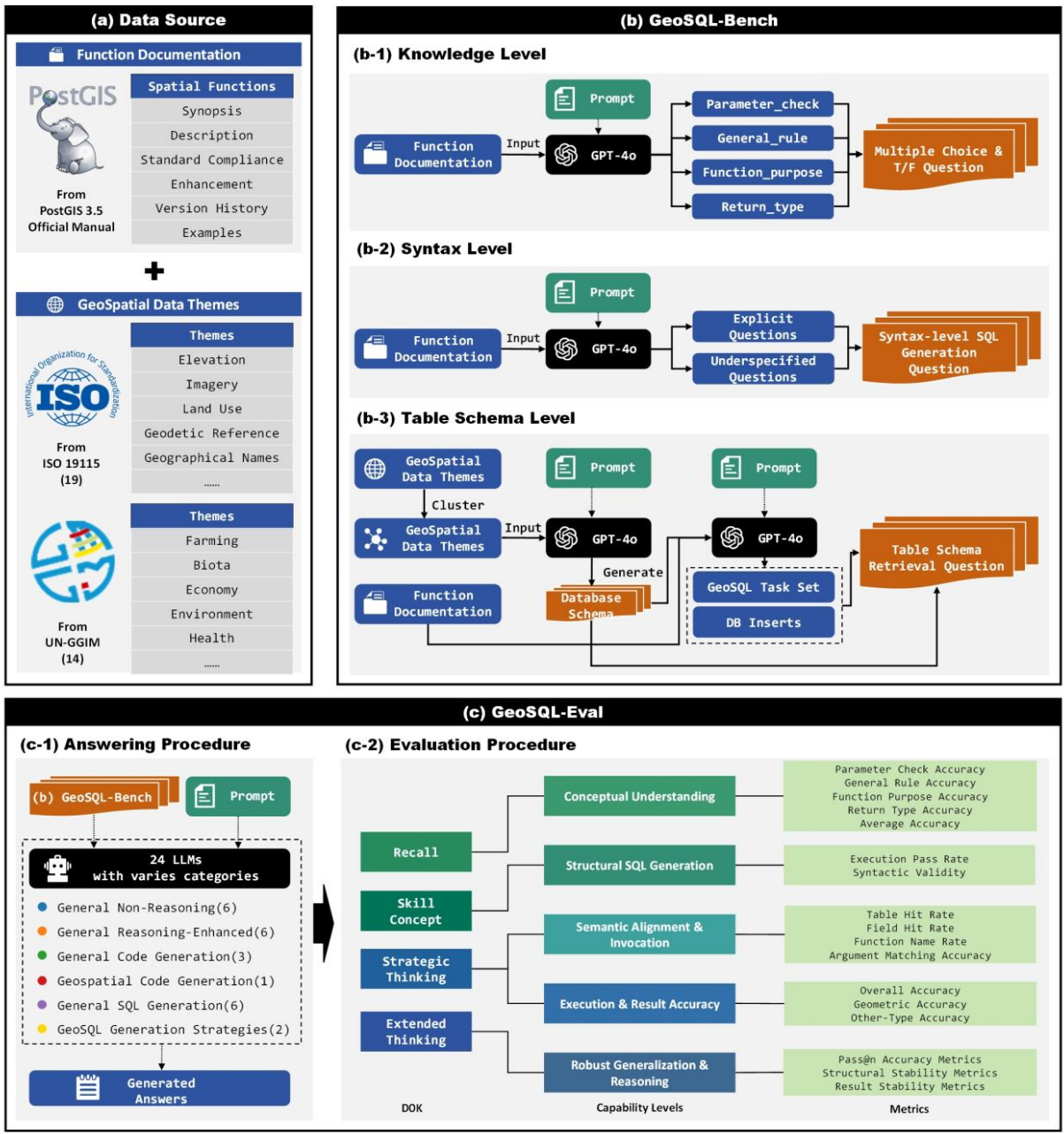

**Figure 4** Diagram of the article process.

- **Conceptual Understanding Layer** evaluates a model's mastery of fundamental knowledge in spatial queries, including function purpose recognition, parameter order and type, return type identification, and comprehension of behavioral specifications such as version dependencies, standard compliance, default settings, and dimensionality preservation strategies.

- **Structured SQL Generation Layer** examines the structural and syntactic correctness of generated queries, encompassing abstract syntax tree (AST) validity checks as well as execution verification within the PostGIS engine.

- **Semantic Alignment & Invocation Layer** focuses on the semantic consistency between generated queries and the database schema, measured by metrics such as table hit rate, field hit rate, function name hit rate, and parameter matching accuracy.

- **Execution & Result Accuracy Layer** validates the effectiveness of model-generated queries from the perspective of execution outcomes, covering accuracy across three representative query types: numerical, Boolean/textual, and geometric.

- **Robust Generalization & Reasoning** assesses model stability and generalization under conditions such as multi-turn interactions, minor input perturbations, or semantic reformulations, with attention to consistency in both structural outputs and final results.

In terms of task design, **GeoSQL-Bench** comprises three categories of tasks, each targeting different levels of capability assessment. The first category consists of **Multiple Choice and True/False Questions**, corresponding to the Conceptual Understanding layer. These tasks focus on evaluating models' comprehension of fundamental spatial query knowledge, including function recognition, awareness of parameter order and type, return type identification, and understanding of behavioral specifications. The second category is **Syntax-level SQL Generation Questions**, constructed from function examples in the PostGIS 3.5 official documentation. These tasks are independent of specific database schemas and are primarily designed to assess the syntactic validity and structural correctness of generated GeoSQL queries. The third category is **Table Schema Retrieval Questions**, built upon the geospatial data themes proposed by UN-GGIM and the MD_TopicCategoryCode classification system in ISO 19115-1:2014. This category encompasses 8 themes and 82 PostGIS geospatial databases, with tasks designed to evaluate models' comprehensive abilities in schema linking and query execution. Notably, the second and third categories jointly support the assessment of all capability levels beyond basic conceptual understanding. Through its multidimensional and hierarchical task system, GeoSQL-Bench provides a comprehensive evaluation of LLMs in GeoSQL query generation, capturing knowledge acquisition, syntactic construction, semantic alignment, execution accuracy, and cross-formulation generalization. This design effectively mitigates risks of overestimating performance due to "accidental correctness" or underestimating utility because of minor structural errors, thereby offering interpretable benchmarks for users from diverse backgrounds in capability analysis, accuracy comparison, and iterative optimization.

The main contributions of this study are as follows:

- We propose the NL2GeoSQL task paradigm, which introduces the evaluation of LLMs on PostGIS query generation, thereby extending the application boundary of NL2SQL.

- We construct **GeoSQL-Bench**, a benchmark consisting of 14,178 tasks across three categories: 2,380 multiple-choice and true/false questions, 3,744 syntax-level SQL generation questions with corresponding distractor formulations, and 2,155 table schema retrieval questions with their distractor counterparts.

- We design and implement **GeoSQL-Eval**, an end-to-end automated evaluation framework covering four cognitive dimensions, five capability levels, and twenty task types. Using this framework, we systematically evaluate 24 models across six categories: General Non-Reasoning Models, General Reasoning-Enhanced Models, General Code Generation Models, Geospatial Code Generation Models, General SQL Generation Models, and GeoSQL Generation Strategy Models.

- Based on the evaluation results, we employ statistical methods such as the entropy weight method, coefficient of variation, skewness, and kurtosis. We further integrate full-process capability coverage—including knowledge acquisition, syntax generation, schema linking, execution accuracy, and robustness—supplemented with analyses of resource consumption and error distributions. This enables in-depth visualization and provides quantitative evidence and concrete recommendations for model improvement and optimization.

- We develop and publicly release the **GeoSQL-Eval leaderboard platform**, which allows research teams worldwide to submit models for testing, supports continuous updates, and fosters cross-team collaboration and knowledge exchange.

The remainder of this study is organized as follows. **Section 2** presents the design rationale, construction methodology, and outcomes of the GeoSQL-Bench benchmark. **Section 3** describes the evaluation methodology, selected models, experimental settings, and performance metrics. **Section 4** reports and analyzes the experimental results. Finally, **Section 5** concludes the study by summarizing the findings, highlighting strengths and limitations, and outlining future research directions.

## 2. GeoSQL-Bench

GeoSQL-Bench comprises three categories of tasks: **Multiple Choice & True/False Questions**, **Syntax-level SQL Generation Questions**, and **Table Schema Retrieval Questions**. This chapter introduces the construction methodology and results for each category in detail.

### 2.1. Data Source

### 2.1.1. PostGIS 3.5 Official Manual

The data used for the first and second categories of tasks is primarily derived from the PostGIS 3.5 official manual. This manual systematically covers core modules including Introduction, Installation and Setup, Data Types, Spatial Functions and Operators, Spatial Indexing, Topology Support, and Performance and Limitations, serving as the authoritative reference for PostGIS users and developers. In this study, the Spatial Functions and Operators section is adopted as the primary basis for task construction, as it provides the most detailed and systematic technical descriptions of PostGIS functions, comprehensively covering their functionalities and usage specifications. An example function is illustrated in **Figure 5**. Let the set of PostGIS functions be defined as:

$$\mathcal{F} = \{f_1, f_2, \dots, f_n\} \tag{1}$$

where $n = 340$ represents the total number of functions documented in the PostGIS 3.5 official manual. Each function $f_i \in F$ can be formalized as a structural 5-tuple:

$$f_i = \langle Sig_i, Desc_i, Std_i, Hist_i, Ex_i \rangle \tag{2}$$

where: $Sig_i$: Synopsis, defining parameters, return type, and calling format; $Desc_i$: Description, specifying the spatial semantics and computational logic; $Std_i$: Standard compliance (e.g., OGC SF, SQL/MM); $Hist_i$: Enhancement and version history; $Ex_i$: A set of usage examples, formally defined as:

$$Ex_i = (q_{i1}, r_{i1}), (q_{i2}, r_{i2}), \dots, (q_{ik}, r_{ik}) \tag{3}$$

where $q_{ij}$ denotes the j-th SQL query example and $r_{ij}$ is the corresponding output, which may take the form of text, graphical representation, or geometric visualization. In total, all 340 functions documented in the PostGIS manual were collected. Since most functions include multiple examples, we ultimately obtained 756 high-quality function example entries.

**Figure 5** Function Documentation Examples

### 2.1.2. UN-GGIM & ISO 19115

The construction of data source themes for the third category of questions is grounded in two authoritative frameworks: the thematic structure of geospatial data proposed by the United Nations Committee of Experts on Global Geospatial Information Management (UN-GGIM) and the MD_TopicCategoryCode classification system defined in ISO 19115-1:2014[35-37]. The former comprises 14 fundamental thematic domains, such as Geodetic Reference, Elevation, Imagery, Geographical Names, Land Use, Population Distribution, and Soil. The latter specifies 19 general topic codes—including farming, biota, economy, elevation, environment, geoscientificInformation, and health—which serve as standardized references for semantic annotation and metadata categorization.

### 2.2. Construction Method

To construct a high-quality evaluation dataset, this study adopts a Self-Instruct data augmentation paradigm based on LLMs, supplemented by systematic revisions informed by expert knowledge. Self-Instruct has become a mainstream approach in both training and data construction for LLMs. Its core mechanism involves providing a small set of examples and explicit instruction templates, thereby guiding the model to generate target-task samples with consistent structure and semantics under contextual references[38-40]. This significantly enhances the efficiency and coverage of sample generation. The generation model employed in this study is GPT-4o released by OpenAI. Although GPT-4o is not the most recent model—by August 2025, GPT-4.1 had already become the widely accessible version with superior performance—it was deliberately selected for the generation phase only. This design effectively prevents potential role overlap between dataset constructors and evaluation participants, ensuring fairness in the construction of test tasks. Moreover, GPT-4o's tasks were confined to linguistic organization of questions and options, format standardization, and sample generalization. All generated content relied strictly on human-provided seed samples, rule-based templates, and structural constraints. As such, its function in the dataset generation process is better understood as a formatting tool rather than a knowledge source, thereby minimizing semantic risk and mitigating the possibility of knowledge leakage[41].

Building upon automated generation, the expert revision phase systematically ensured the quality of the questions[42]. For multiple-choice questions, experts evaluated whether each item had a single correct answer and confirmed that the distractors included at least one plausible option, thereby preventing models from inferring the correct answer solely through semantic cues. For true/false questions, the focus was on verifying that each item corresponded to a clear and unambiguous knowledge point, while eliminating statements that could be reasonably interpreted as either correct or incorrect. In SQL generation tasks, experts executed the reference answers to verify syntactic correctness, assess the reasonableness of the output, and ensure consistency between referenced fields and the database schema, thus maintaining rigor and reproducibility at the execution level. To enhance objectivity and consistency, three experts with backgrounds in surveying, geographic information science, and data modeling were invited to participate in the review process. They were designated as Expert 1, Expert 2, and Expert 3 in descending order of seniority, as shown in **Table 1**. The review followed a "dual-review and single-adjudication" mechanism: Experts 2 and 3 independently evaluated and revised the model's draft, after which Expert 1 synthesized and finalized their feedback into the definitive version of each question. This three-stage joint review process not only preserved the efficiency of question generation but also ensured accuracy and semantic consistency, thereby providing institutional safeguards for the scientific validity of the benchmark. The detailed construction methods for each question type are presented in the following sections.

Table 1 Background and Selection Criteria of Evaluation Experts.

| No. | Age | Qualification | Research Field | Selection Criteria |
|---|---|---|---|---|
| Expert 1 | 59 | Professor, PhD Advisor | Surveying and Geoinformatics | Extensive experience in national key research projects and geospatial modeling. Provides overall judgment to ensure the scientific rigor and validity of the test items. |
| Expert 2 | 40 | Associate Professor | Geographic Information Science | Rich academic background in GIS and spatial databases. Responsible for independently reviewing the logical consistency and semantic clarity of the test items. |
| Expert 3 | 26 | PhD | Data Modeling and Spatial Analysis | Strong expertise in data modeling and SQL implementation. Verifies the executability and reproducibility of the test items within the database framework. |

**2.2.1. Multiple Choice & T/F Question**

The multiple-choice and true/false question formats encompass four categories of cognitive tasks, all constructed on the basis of the PostGIS 3.5 Official Manual as outlined in **Section 2.1.1**. These tasks are designed to systematically assess LLMs' fundamental understanding and semantic comprehension of spatial database functions. The first category, **Function Purpose Recognition** (multiple-choice), evaluates the model's ability to identify the core functionality of PostGIS functions. The question types include "select the function name based on its description" and "determine the functionality based on the function name." These items are derived from the functional description section ($Desc_i$) of the documentation. Three questions were generated for each function, yielding a total of 680 items. The second category, **Parameter Check Recognition** (true/false), focuses on the model's knowledge of parameter order and type matching, testing whether it can accurately identify the type, order, and number of input parameters. These questions were constructed from the function's synopsis ($Sig_i$), with two items per function, resulting in 680 questions. The third category, **Return Type Recognition** (multiple-choice), measures the model's ability to identify the output data type of each function, also based on $Sig_i$. Each function corresponds to one item, producing 340 questions in total. The fourth category, **General Rule Compliance** (true/false), assesses the model's understanding of behavioral constraints and version compatibility of functions. The content covers aspects such as compliance with standards (e.g., OGC, SQL/MM), default behaviors (e.g., automatic deduplication, preservation of Z-values), output structure properties (e.g., dimensional retention, geometry support), and version evolution. These items are derived from the standards and historical notes ($Std_i$ and $Hist_i$), with two questions per function, totaling 680 questions. All four categories were developed with explicit construction rules, reference materials, and instruction templates. The sources of reference content, the design logic of prompts, and representative question examples are illustrated in **Figure 6**.

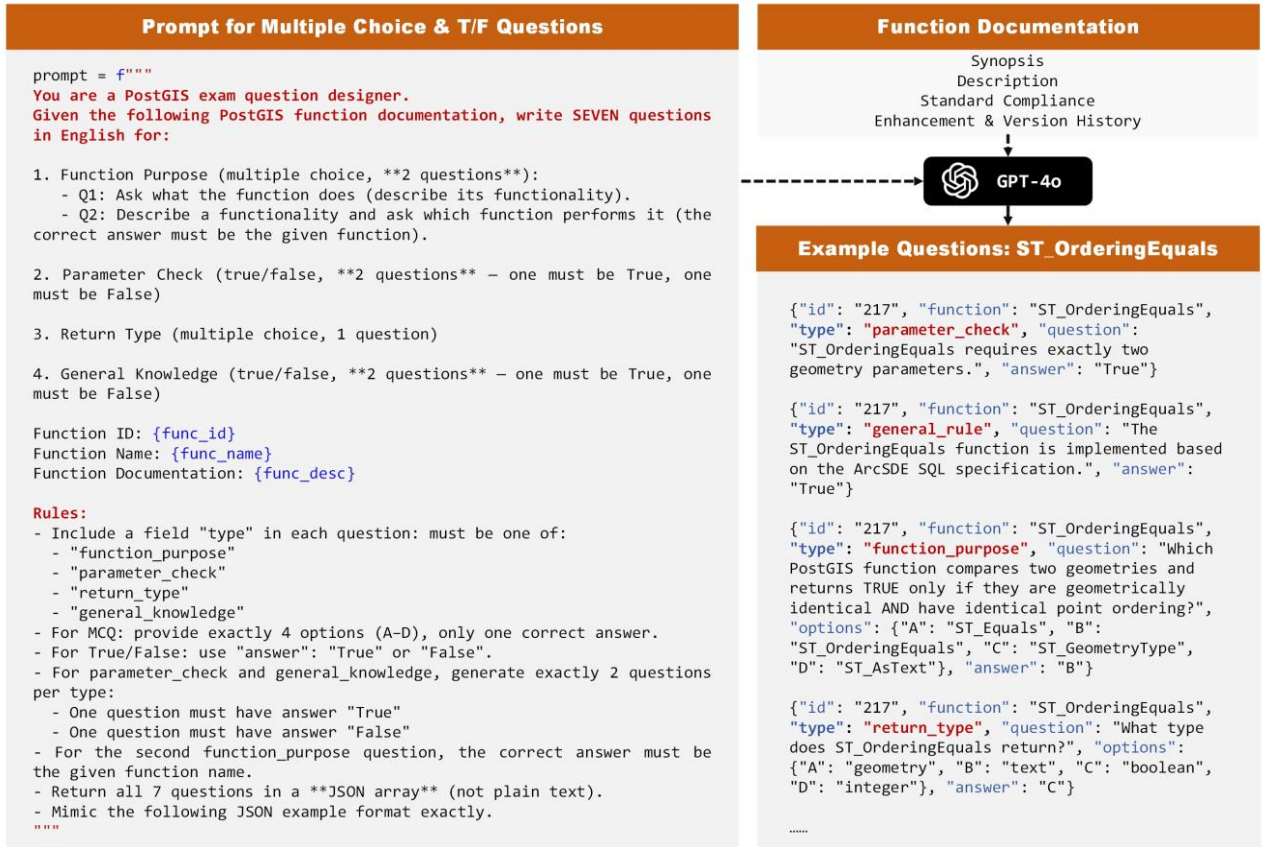

**Figure 6** Multiple Choice & T/F Question Construction Framework

**2.2.2. Syntax-level SQL Generation Question**

The syntax generation tasks are designed to evaluate whether LLMs can, based on natural language requirements, correctly select appropriate functions and generate PostGIS queries that are semantically complete, syntactically valid, and logically executable. The construction of these tasks relies on the function's synopses ($Sig_i$), functional descriptions ($Desc_i$), and usage examples ($Ex_i$) provided in the PostGIS 3.5 Official Manual, as described in **Section 2.1.1**. By prompting GPT-4o to restructure these examples, two variants of natural language inputs were created for each function example. The first variant, termed the **Explicit Prompt**, contains clear and complete semantic information, including the query objective, input data, and expected output type. This form simulates standardized queries posed by professional users. The second variant, termed the **Underspecified Prompt**, deliberately weakens the input by introducing terminological ambiguity, omitting elements of query intent, or altering sentence structure. This design simulates vague queries commonly encountered from non-expert users or in real-world application contexts. For instance, an underspecified prompt may omit the spatial reference system (SRID), confuse the order of function parameters, simplify the specification of geometry types, or replace technical terminology with metaphorical language (e.g., using "draw a circle around it" instead of "buffer query"), thereby serving as a robustness test.

All natural language inputs were generated by GPT-4o, from which the system automatically extracted structured information such as function names, parameter order, and parameter types to construct candidate questions. The generated SQL statements were then executed in a real PostGIS environment to obtain the corresponding query results. Human experts reviewed each question to verify that the natural language description was consistent with the target function in terms of logic, syntax, and parameters, thereby ensuring both executability and correctness of the results. Ultimately, each syntax generation task was formally represented as the following sextuple:

$$Q_i^{gen} = \langle Func_i, Args_i, NL_i, Ex_i, Ans_i \rangle \qquad (4)$$

where: $Func_i$ denotes the function name involved; $Args_i$ specifies the parameter types and their order; $NL_i$ is the natural language description, which serves as the input for model responses; $Ex_i$ represents the PostGIS query example; and $Ans_i$ provides the execution result, serving as the reference answer against which subsequent outputs are compared. The detailed strategy for description generation, prompt design, and representative construction examples is illustrated in **Figure 7**.

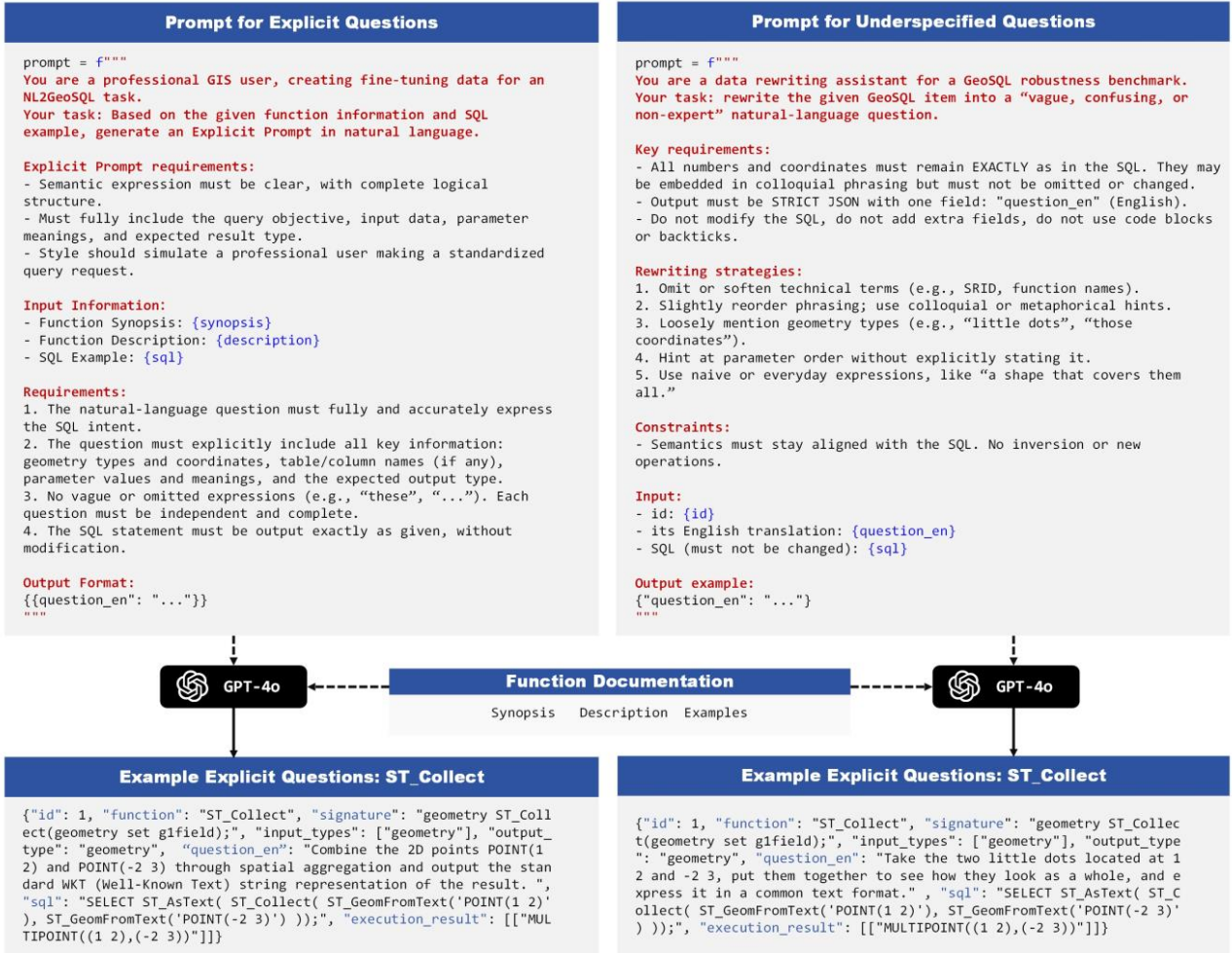

Figure 7 Syntax-level SQL Generation Question Construction Framework

**2.2.3. Table Schema Retrieval Question**

Because the table schema retrieval questions require models not only to interpret natural language intent but also to accurately identify and utilize the relevant table schema before generating SQL statements, their cognitive complexity exceeds that of conventional syntax generation tasks. To systematically construct this category of questions, the process was divided into two stages: database construction and database-driven SQL question design.

In the database construction stage, the semantic basis for schema design was derived from the thematic framework of the United Nations Committee of Experts on UN-GGIM and the MD_TopicCategoryCode classification defined in ISO 19115-1:2014, as discussed in **Section 2.1.2**. To aggregate core themes and generate the database schema, a hierarchical clustering approach was employed in combination with a function-driven data generation strategy. Specifically, pretrained Word2Vec embeddings were used to vectorize theme descriptions from both classification systems. A 300-dimensional model trained on the Google News corpus was adopted, converting tokenized and stopword-filtered descriptions into semantically rich vector representations[43]. Pairwise Euclidean distance matrices were then computed to quantify semantic dissimilarities, which served as input for hierarchical clustering using Ward's linkage method. This method iteratively merged themes by minimizing within-cluster variance, producing a dendrogram with a hierarchical structure. The quality of clustering across different theme counts was evaluated using the silhouette coefficient, which indicated that dividing the dataset into eight clusters yielded the best performance. Finally, the clustering results were refined and interpreted by experts, resulting in eight core database themes, as summarized in **Table 2**.

**Table 2 Clustering Results and Correspondence to UN-GGIM Themes and ISO 19115-1 Topic Categories.** The "Category" column represents the names assigned by experts based on the clustering results. Each category includes two themes, with their corresponding sub-themes shown under the "UN-GGIM Themes" and "ISO 19115-1 Topic Categories" columns.

| ID | Category | UN-GGIM Themes | ISO 19115-1 Topic Categories |
|---|---|---|---|
| 1 | Natural Resources & Ecosystems | Geology & Soils | Farming; Biota; Geoscientific Info |
| 2 | Natural Environment & Climate Change | Elevation & Depth; Water | Climate & Atmosphere; Elevation; Inland Waters; Oceans |
| 3 | Urban Planning & Land Use | Land Cover/Use; Land Parcels; Functional Areas | Boundaries; Planning & Cadastre |
| 4 | Buildings, Facilities & Infrastructure | Buildings & Settlements; Physical Infrastructure; Transport Networks | Structures; Transportation; Utilities & Communication |
| 5 | Population & Social Services | Population Distribution | Society; Economy |
| 6 | Environmental Protection & Disaster Response | —— | Environment; Disaster; Intelligence/Military |
| 7 | Remote Sensing & Spatial Observation | Orthoimagery | Imagery/Base Maps/Earth Cover; Extra-Terrestrial |
| 8 | Geographic Names & Spatial Cognition | GGRF; Geographic Names; Addresses | Location |

Building on this foundation, a function-driven data generation strategy is employed to ensure broad adaptability of the generated database structures to diverse functions. Specifically, four to eight spatial functions are randomly selected from the PostGIS function set, and their function's synopses ($Sig_i$), descriptions ($Desc_i$), and usage examples ($Ex_i$) are extracted. These elements are then bound to predefined thematic contexts and used as prompts to guide GPT-4o in generating database structures that reflect authentic business semantics. The generated databases must satisfy several critical constraints: (1) the structural design must strictly align with the specified theme to prevent semantic drift; (2) all fields must comply with the input and output type requirements of the associated PostGIS functions; (3) each table must contain at least one spatial attribute and provide realistic English sample values to ensure executability; and (4) field names must remain contextually relevant, avoiding template-based expressions or default model-generated vocabulary. This strategy aims to simultaneously guarantee the operability of functions and the semantic consistency of themes. It ensures that each function is covered at least once, with each database consisting of one or more tables. Following this approach, a total of 82 databases were constructed. Let the structure of the generated database be defined as follows:

$$\mathcal{D} = \langle Name, Topic, \mathcal{T} \rangle \tag{5}$$

where: "*Name*": denotes the database name, which must carry realistic business semantics and avoid template-style naming (e.g., "UrbanMonitoringDB," "GeoTrack"); "*Topic*": represents the thematic focus of the database, derived from the predefined topic set introduced in **Section 2.1.2**; $\mathcal{T} = \{T_1, T_2, \ldots, T_n\}$: the set of tables contained in the database, where $n \geq 1$. Each database must include at least one business-relevant entity table with a non-trivial schema capable of supporting spatial queries, table joins, and other relational operations. Each table $T_i \in \mathcal{T}$ is defined as a sextuple:

$$T_i = \langle TableName_i, Schema_i, PK_i, FK_i, Geo_i, Samples_i \rangle \tag{6}$$

with the following components: $TableName_i$: the table name, which must convey clear business semantics. $Schema_i = \{(f_1, \tau_1), (f_2, \tau_2), \ldots, (f_k, \tau_k)\}$: a set of field–type pairs, where the number of fields satisfies $k \in [3,5]$, The field types $\tau_j \in T_{PostgreSQL}$, such as VARCHAR, INTEGER, DATE, BOOLEAN, DECIMAL, and geometry (e.g., POINT, 4326). $PK_i \subseteq \{f_1, \ldots, f_k\}$: the set of primary key fields. $FK_i \subseteq \{f_1, \ldots, f_k\}$: the set of foreign key fields. $Geo_i \subseteq \{f_1, \ldots, f_k\}$: the set of spatial fields (e.g., geometry type attributes). Each table must include at least one spatial field. $Samples_i = \{r_1, r_2\}$: a set of sample tuples, where each $r_j$ is a complete record with values conforming to their respective data types. The samples must be expressed in English or WKT format. For example: $r_1$ = (101, "JohnDoe", "2010-08-23", 50000.00, 1, "POINT(120.3836.06)").

After the database and table structures are generated, the system proceeds to the construction stage of table schema retrieval questions. During initial design, each database is bound to 4–8 PostGIS functions. This stage centers on the bound functions, generating corresponding tasks and their associated data insertion statements one by one. Each task is constructed around a target function, requiring that the natural language query explicitly reflect the semantic parameters of that function, while optionally incorporating other bound functions within the same database to enhance semantic complexity and coverage. In the generation process, the system provides GPT-4o with the relevant table metadata, along with the function's synopsis ($Sig_i$), description ($Desc_i$), and usage examples ($Ex_i$). When multiple tables are involved, metadata from all relevant tables is supplied simultaneously. With this structural and functional information, the model is prompted to generate natural language query tasks with clear semantic objectives, and to automatically infer the dependent table names, field names, and function calls, thereby constructing structured task tuples and corresponding data insertion statements. The insertion statements are required to differ from the provided examples to avoid redundancy across tasks; subsequently, these statements are used for batch data generation and undergo expert manual review. All generated SQL queries must pass runtime verification to ensure they return valid, non-empty results. The table schema retrieval question is formally extended as a eight-tuple:

$$Q_i^{schema} = \langle Func_i, Args_i, NL_i, Sql_i, Database_i, Schema_i, Samples_i, Data_i \rangle \tag{7}$$

The five additional components are defined as follows: $Sql_i$: the standard GeoSQL query corresponding to the user's natural language question; $Database_i$: the name of the target database; $Schema_i$: the structural definition of the relevant tables, including table names, field names, types, primary and foreign keys, spatial fields, and two representative sample records; $Samples_i$: sample data drawn from the corresponding tables; $Data_i$: the constructed insertion SQL statements ensure that $Sql_i$ produces non-empty query results.

For this type of task, we first construct **Explicit Prompts**, in which the key information in the natural language request directly corresponds to the field names in the table schema, requiring no additional reasoning for resolution. Building on this, we further extend the task design with **Underspecified Prompts**. In such cases, discrepancies or semantic shifts exist between the natural language expressions and the actual schema fields. For example, the query may reference *"building height"* while the database field is named *"elevation"*. These tasks are designed to evaluate whether a model can achieve field-level semantic alignment by reasoning over and mapping schema information under conditions of semantic inconsistency. The construction methodology of table schema retrieval questions and the primary prompt templates used in this process are illustrated in **Figure 8**. Examples of the generated database schemas, GeoSQL Q&A tasks, and insertion statements are shown in **Figure 9**. A complete list of thematic databases, including the number of tables and fields within each, is provided in **Appendix A, Table A1**.

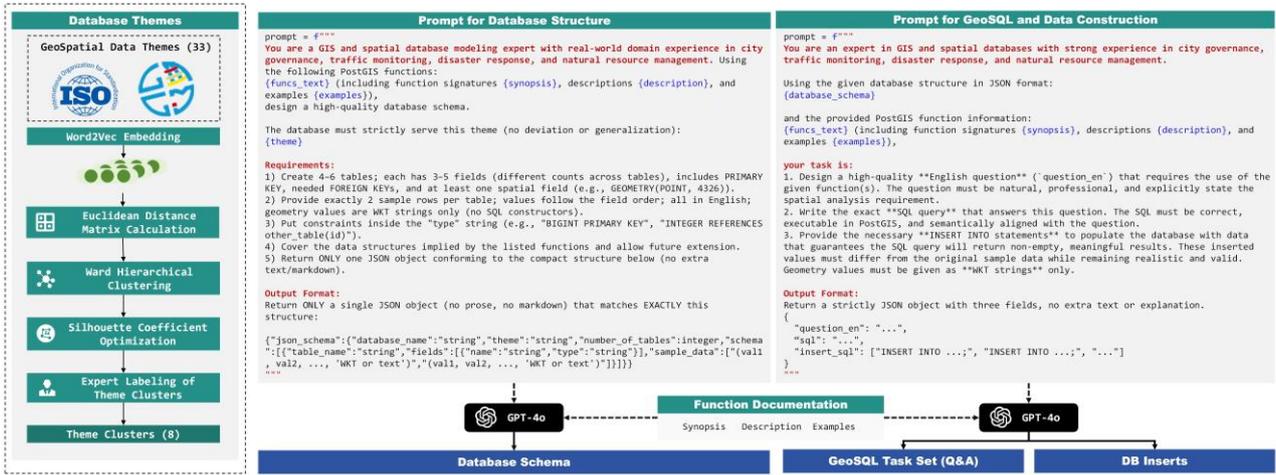

**Figure 8** Table schema retrieval question construction framework

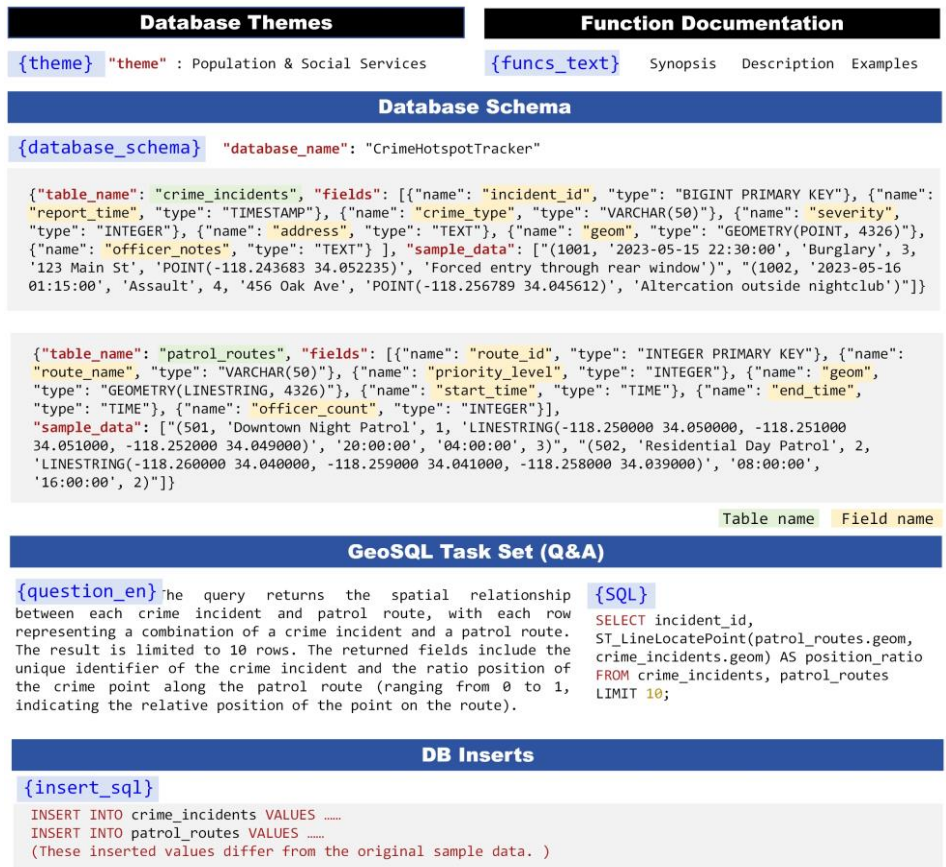

**Figure 9** Example of table schema retrieval question

## 2.3. Construction Result

Finally, the number of tasks generated in GeoSQL-Bench and their distribution across the GeoSQL-Eval evaluation framework are summarized in **Table 3**, demonstrating comprehensive coverage of all PostGIS functions.

**Table 3** Distribution of GeoSQL-Bench Question Types across the GeoSQL-Eval Framework

| Major Question Type | Subtask / Category | Number of Questions | Covered Capability Levels in GeoSQL-Eval |
|---|---|---|---|
| **Multiple Choice & T/F Question** | Function Understanding | 680 | Conceptual Understanding |
| | Parameter Matching | 680 | |
| | Return Type Recognition | 340 | |
| | Behavioral Norms | 680 | |

| Major Question Type | Subtask / Category | Number of Questions | Covered Capability Levels in GeoSQL-Eval |
|---|---|---|---|
| | Subtotal | 2380 | |
| **Syntax-level SQL Generation Question** | Syntax-valid SQL Generation | 3744 | Other Four Capability Levels |
| | Underspecified SQL Generation | 3744 | |
| | **Subtotal** | **7488** | |
| **Table Schema Retrieval Question** | Explicit Prompt | 2155 | Other Four Capability Levels |
| | Underspecified Prompt | 2155 | |
| | **Subtotal** | **4310** | |
| **Total** | — | **14,178** | **Full Coverage of Five Capability Levels** |

## 3. GeoSQL-Eval

The GeoSQL-Eval framework is structured around the five capability levels and twenty evaluation task categories illustrated in **Figure 4**, encompassing both the answering and judging processes for the evaluated models. This chapter introduces the answering workflow, judging workflow, and evaluation metrics of the GeoSQL-Eval framework.

### 3.1. Answering Procedure

The answering process for all models is controlled through prompt guidance. Within the Conceptual Understanding layer, the evaluation tasks consist of multiple-choice and true/false questions. A unified standardized prompt constrains the models to respond solely based on the question stem, prohibiting any explanatory text and allowing only a single output option: multiple-choice questions must be answered with "A," "B," "C," or "D," and true/false questions with "True" or "False." The model outputs are stored in **.jsonl** format, with each line corresponding to the answer of one question, enabling streamlined line-by-line reading and evaluation. Examples of the prompt formats are provided in **Figure 10 (a) and (b)**. The answers are denoted as $Objective\_Answer_{i,j}$, where $i$ represents the index of the evaluation model (Model Index) and $j$ represents the index of the evaluation question (Question Index); thus, $Objective\_Answer_{i,j}$ denotes the answer of the $i$-th model to the $j$-th question.

For the remaining capability levels, evaluation is conducted on the basis of the PostGIS queries generated by the models and their corresponding execution results. To ensure fairness and consistency while minimizing prompt-induced interference with model behavior, the system employs concise natural language instructions to guide responses. Specifically, for syntax-level SQL generation questions, the model generates the corresponding SQL query *solely* from the natural language description $NL_i$, without providing any explanatory text. For table schema retrieval questions, the model must generate SQL queries based on the natural language description $NL_i$, the schema information $Schema_i$, and representative sample data, again without additional explanations. All query outputs are stored in .jsonl format, recorded sequentially by question index to facilitate programmatic iteration and evaluation. Examples of the prompt formats are shown in Figure 10(c) and (d). Each generated query is then executed in a real PostGIS environment: if execution succeeds, its execution state is marked as 1 and both the query result and its type (text, Boolean, numeric, or geometry) are recorded; if execution fails, the execution state is marked as 0 and the result is empty. In addition, syntax validation is performed using the Python-pglast library, an AST parser for PostgreSQL: successful parsing is marked as 1 and the syntax tree is saved, while parsing failures are marked as 0. Consequently, for each GeoSQL task, the model's output is evaluated across the following five dimensions:

$$GeoSQL\_Answer_{i,j} = \langle ExecState_{i,j}, ExecResult_{i,j}, ResultType_{i,j}, SyntaxState_{i,j}, SyntaxTree_{i,j} \rangle \quad (8)$$

Similarly, $i$ denotes the $i$-th evaluation model (Model Index), and $j$ denotes the $j$-th evaluation question (Question Index). Thus, $GeoSQL\_Answer_{i,j}$ represents the answer of the $i$-th model to the $j$-th question. To assess the stability of generation results and identify potential hallucination phenomena, each task type is executed five independent times for every model under a unified prompt setting. All outputs from these runs are recorded, thereby providing the empirical basis for subsequent analyses of hallucination and the stability of generation quality.

**Figure 10** Prompt templates used for different question types.

### 3.2. Evaluation Procedure

The evaluation process is organized around twenty tasks spanning five capability levels; accordingly, this study introduces the computational methods for each capability level on a level-by-level basis.

#### 3.2.1. Conceptual Understanding Layer

The scoring method is based on strict answer matching, employing a rule-based mechanism to compare model outputs against the reference answers, where only exact matches are considered correct. The performance metric for this capability level is denoted as $Accuracy_{Foundational}$.

$$Accuracy_{Foundational} = \frac{1}{N} \sum_{i=1}^{N} 1(\hat{a}_i = a_i) \quad (9)$$

Here, $N$ represents the total number of evaluation tasks within the capability level; $\hat{a}_i$ denotes the model's output for the $i$-th question; $a_i$: is the corresponding reference answer; and $1(\cdot)$ is an indicator function that equals 1 if the expression inside is true, and 0 otherwise.

#### 3.2.2. Structured SQL Generation Layer

In this layer, the evaluation primarily focuses on the executability and syntactic validity of generated queries. Two core metrics are defined: **Execution Pass Rate** and **Syntax Accuracy**. The Execution Pass Rate measures the proportion of model-generated queries that can be successfully executed in a real PostGIS environment. It is defined as the ratio of queries satisfying $ExecState_{i,j} = 1$ to the total number of generated queries, as expressed by the following equation:

$$EPR = \frac{1}{N}\sum_{i=1}^{N} \mathbb{I}(ExecState_{i,j} = 1) \qquad (10)$$

**Syntax Accuracy (SA)** evaluates whether the generated SQL statements conform to PostgreSQL syntax specifications. The Python-pglast library is employed to parse the AST; a query is considered syntactically correct if parsing succeeds. It is defined as the proportion of statements satisfying $SyntaxState_{i,j} = 1$ over the total number of generated queries, as expressed by the following equation:

$$SA = \frac{1}{N}\sum_{i=1}^{N} \mathbb{I}(SyntaxState_{i,j} = 1) \qquad (11)$$

Here, $N$ denotes the total number of tasks, and $\mathbb{I}(\cdot)$ is the indicator function.

### 3.2.3. Semantic Alignment and Invocation Layer

In this layer, the primary focus is whether the PostGIS queries generated by LLMs achieve accurate semantic alignment with the intended function. To this end, four key performance metrics are defined: **Table Hit Rate**, **Field Hit Rate**, **Function Name Rate**, and **Argument Matching Accuracy**. Prior to evaluation, each model-generated query $GeoSQL\_Answer_{i,j}$ must undergo structured parsing to extract core elements, including table names, field names, function names, and parameter order and types. This parsing is performed on the query's abstract syntax tree $SyntaxTree_{i,j}$, where a traversal algorithm extracts all structural information and standardizes it into JSON format, facilitating alignment with the reference answers. The keywords within the syntax tree exhibit stable structural features: table names are identified by the *RangeVar.relname* node; field names are given by the *ColumnRef.fields* node; parameter order and types are derived from *FuncCall.args*; and function names are provided by the *FuncCall.funcname* node. It is important to note that each syntax-level SQL generation question is constructed from a single standard function example provided in the official PostGIS documentation (**see Section 2.2.2**), while each table schema retrieval question also revolves around one primary function, with other functions included only as auxiliary or semantic complements (see **Section 2.2.3**). Consequently, each task essentially involves a single core function, and the evaluation therefore prioritizes verifying whether the model correctly invokes this core function, followed by checking whether its parameter types and order are consistent with the reference answer. To efficiently accomplish the above structural extraction, we design a unified syntax tree traversal algorithm based on depth-first search (DFS). The algorithm recursively traverses all syntax tree nodes, captures SQL-relevant structural elements, and encapsulates them into standardized JSON dictionaries, thereby supporting subsequent metric computation and error analysis. Theoretically, the algorithm achieves both time and space complexity of *O(N)*, where *N* denotes the total number of syntax tree nodes. The core pseudocode of this module is provided in **Table 4**.

Table 4 Core Pseudocode of the Unified Syntax Tree Traversal Algorithm

| Algorithm: ExtractPostGISStructure(SQL_Query) |
|---|
| **Input:** |
|     SQL_Query - A SQL string containing a PostGIS query |
| **Output:** |
|     Structure_Info - A dictionary containing: |
|         - Table names |
|         - Column names |
|         - Function names |
|         - Function argument expressions |

---

**Algorithm: ExtractPostGISStructure(SQL_Query)**

**Procedure:**
01  Parse SQL_Query into AST
02  tree ← parse_sql(SQL_Query)

03  Initialize:
04      Tables ← ∅
05      Columns ← ∅
06      Functions ← []
07      FunctionArgs ← []

08  Define recursive procedure Visit(Node):
09      If Node is a TableReference:
10          Tables ← Tables ∪ {Node.relname}

11      If Node is a ColumnReference:
12          col_name ← Join(Node.fields, '.')
13          Columns ← Columns ∪ {col_name}

14      If Node is a FunctionCall:
15          func_name ← Join(Node.funcname, '.')
16          Functions.append(func_name)

17          args ← []
18          For each arg in Node.args:
19              raw ← RawSQL(arg)
20              args.append(raw)
21          FunctionArgs.append({func_name: args})

22      For each child in Node.children:
23          Visit(child)

24  Visit(tree.root)

25  Return {
26      "tables": Tables,
27      "columns": Columns,
28      "functions": Functions,
29      "function_args": FunctionArgs
30  }

---

Therefore, after preprocessing with the traversal algorithm, the system parses the $SyntaxTree_{i,j}$ of each model-generated $GeoSQL\_Answer_{i,j}$ to extract the following four categories of structural information for comparison with the reference answers:

$$ExtractedStruct_{i,j} = \{E\_Tables_{i,j}, E\_Fields_{i,j}, E\_FuncName_{i,j}, E\_FuncArgs_{i,j}\} \qquad (12)$$

Here, $E\_Tables_{i,j}$ and $E\_Fields_{i,j}$ denote the extracted sets of table and field names, which are present only in table schema retrieval questions. $E\_FuncName_{i,j}$ represents the core function name invoked in the query, while

$E\_FuncArgs_{i,j}$ refers to the function's parameter information, including their order and corresponding data types. Because table schema retrieval questions often involve table and field renaming, precise parameter matching cannot be guaranteed; therefore, this element is only applicable to syntax-level SQL generation questions.

Based on the above structural extraction, four evaluation metrics are defined. The **Table Hit Rate (THR)** measures whether the standard table names $Table_i \in Q_i^{schema}$ are correctly identified and appear in the extracted table set $E\_Tables_{i,j}$ from the model output. If none of the identified tables are correct, the score is 0. If correct matches exist, the score is calculated as the percentage of correctly recognized tables relative to the total number of tables used. The formula is defined as:

$$THR_{i,j} = \begin{cases} 0, & \text{if } Table_i \cap E\_Tables_{i,j} = \emptyset \\ \frac{|Table_i \cap E\_Tables_{i,j}|}{|E\_Tables_{i,j}|}, & \text{otherwise} \end{cases} \quad (13)$$

The **Field Hit Rate (FHR)** evaluates the coverage of the standard field set $Fields_i \in Q_i^{schema}$ within the extracted field set $E\_Fields_{i,j}$. If no fields are matched, the score is 0. If matches exist, the score is calculated as the proportion of correctly recognized fields relative to the total number of fields. The formula is defined as:

$$FHR_{i,j} = \begin{cases} 0, & \text{if } Table_i \cap E\_Fields_{i,j} = \emptyset \\ \frac{|Fields_i \cap E\_Fields_{i,j}|}{|E\_Fields_{i,j}|}, & \text{otherwise} \end{cases} \quad (14)$$

The **Function Name Hit Rate (FNR)** evaluates whether the standard function name $Func_i \in Q_i$ is accurately identified within the extracted function name set $E\_FuncName_{i,j}$. If correctly identified, the score is 1; otherwise, it is 0. The formula is defined as:

$$FNR_{i,j} = \begin{cases} 1, & \text{if } Func_i \in E\_FunctionName_{i,j} \\ 0, & \text{otherwise} \end{cases} \quad (15)$$

The **Argument Match Accuracy (AMA)** refines the evaluation by measuring the full-order matching degree of function parameters. Specifically, it compares the parameter type list $E\_FuncArgs_{i,j}$ from the model output with the reference parameter sequence $Args_i$, while preserving parameter order constraints. Let the reference sequence be $Args_i = \{a_1, a_2, ..., a_n\}$ and the model-generated sequence be $E_{FuncArgs i,j} = \{b_1, b_2, ..., b_m\}$. The AMA score is computed as the number of parameters that exactly match in both type and position, divided by the total number of parameters in the reference answer. For instance, if the reference is {geometry, int, bool} and the model output is {geometry, bool, int}, only the first parameter matches, yielding a match count of 1 out of 3, i.e., an accuracy of 33.33%. The formula is given below, where $\mathbb{I}[\cdot]$ denotes the indicator function, returning 1 if the condition holds and 0 otherwise.

$$\text{AMA}_{i,j} = \frac{1}{n} \sum_{k=1}^{n} \mathbb{I}[a_k = b_k] \quad (16)$$

### 3.2.4. Execution and Result Accuracy Layer

This layer is designed to assess the consistency between the SQL queries generated by the model and the reference answers. To account for variations in result data types, tailored comparison strategies and accuracy metrics are employed. Specifically, distinct evaluation methods are established for geometric data types and for numeric, textual, or Boolean types.

#### 3.2.4.1. Geometry type

For the evaluation of geometric types, if both the $Type_i$ specified in the task and the $ResultType_{i,j}$ generated by the model are classified as either geometry or geography, the column is identified as geometric and assessed using the corresponding strategy. To facilitate comparison, geometry types are preserved, while geography types are uniformly converted to geometry. Since the task imposes no restrictions on the output format of geometric objects, the reference answers and the model-generated PostGIS results may appear in WKB, WKT, or EWKT representations, where WKB and EWKT include SRID information, and WKT does not. To ensure consistency, all geometric objects are re-projected to *SRID=4326* and subsequently standardized into both WKB and EWKT formats. During evaluation, the *ST_AsEWKT* function is first employed to compare the EWKT representations of the model output and the reference answer; if they match exactly at the EWKT level, the result is considered correct.

$$C_1: ST\_AsEWKT\left(ExecResult_{i,j}^{(k)}\right) = ST\_AsEWKT\left(Ans_i^{(k)}\right) \tag{17}$$

Otherwise, the comparison proceeds by applying the *ST_Equals(ST_SnapToGrid(geom, 1e-5))* function to the WKB representations, thereby evaluating topological equivalence while allowing a numerical tolerance of 1e-5. For three-dimensional geometries, the *ST_DumpPoints* function is further employed to extract all vertex Z-values for point-by-point comparison. The result is accepted if the differences in Z-values across all corresponding points do not exceed 1e-6, or if both geometries lack Z-values, indicating two-dimensional objects. Only when both topological consistency and Z-value precision requirements are satisfied is the column deemed fully correct. The evaluation formula is as follows:

$$G_{i,j,k}^{exec} = ST\_GeomFromWKB(ExecResult_{i,j}^{(k)}) \tag{18}$$

$$G_{i,k}^{ans} = ST\_GeomFromWKB\left(Ans_i^{(k)}\right) \tag{19}$$

$$\tilde{G}_{i,j,k}^{exec} = ST\_SnapToGrid(G_{i,j,k}^{exec}, 10^{-5}) \tag{20}$$

$$\tilde{G}_{i,j,k}^{anw} = ST\_SnapToGrid(G_{i,j,k}^{anw}, 10^{-5}) \tag{21}$$

$$C_2: ST_{Equals}(\tilde{G}_{i,j,k}^{exec}, \tilde{G}_{i,k}^{ans}) \tag{22}$$

$$C_3 := \max_k \left|Z_k^{(j)} - Z_k^{(a)}\right| \leq 10^{-6} \text{ or both are 2D (Z=None)} \tag{23}$$

$$A_{i,j}^{geom} = \begin{cases} 1, & \text{if } C_1 \\ 1, & \text{if } C_2 \text{ and } C_3 \\ 0, & \text{otherwise} \end{cases} \tag{24}$$

$$Accuracy_{Geom} = \frac{1}{n}\sum_{i=1}^{n} A_{i,j}^{geom} \tag{25}$$

**3.2.4.2. Numerical /Text/Boolean type**

For SQL query results of numeric (e.g., integer, float), textual, or Boolean types, the system adopts a value-matching strategy for evaluation. Specifically, both the model output and the reference answer are first normalized by removing all whitespace characters (spaces, tabs, newlines, etc.) and converting the text to lowercase, thereby ensuring case-insensitive and whitespace-insensitive comparison. If the processed values are identical across all rows, the column is considered correctly matched. The normalization operation is defined as:

$$norm\left(ExecResult_{i,j}^{(k)}\right) := lower\left(removeWS\left(ExecResult_{i,j}^{(k)}\right)\right) \tag{26}$$

Accordingly, the correctness metric for text and Boolean fields is defined as:

$$A_{i,j}^{text} = \begin{cases} 1, & \text{if } \forall k \in \{1,\dots,n\}, norm\left(ResultType_{i,j}^{(k)}\right) = norm\left(Ans_i^{(k)}\right) \\ 0, & \text{otherwise} \end{cases} \tag{27}$$

$$Accuracy_{Num/Text/Boolean} = \frac{1}{n}\sum_{i=1}^{n} A_{i,j}^{text} \tag{28}$$

### 3.2.4.3. Pass@n

As described in **Section 3.1 *Answering Procedure***, each model generates five independent responses for the same question. Building upon the three aforementioned accuracy metrics, this study further introduces the pass@n measure. This metric is defined as the proportion of cases in which at least one of the n generated responses is correct, thereby reflecting the model's success rate under multiple-generation conditions. The calculation formula is given as follows:

$$pass@n = 1 - \frac{C_n}{N} \tag{29}$$

where $N$ denotes the total number of generated samples and $C_n$ represents the number of incorrect samples. To enhance the robustness of GeoSQL query accuracy evaluation, pass@1, pass@3, and pass@5 are adopted in this study.

### 3.2.5. Generalization and Robust Reasoning

This layer aims to provide a comprehensive evaluation of both the stability of model outputs across multiple SQL generation rounds and their robustness under variations in natural language expressions. Output stability reflects whether repeated generations exhibit randomness or "hallucinations," that is, whether the results remain consistent across multiple attempts. As described in **Section 3.1 *Answering Procedure*,** each model produces five independent responses to the same question. To quantify this stability, the coefficient of variation (CV) is adopted as an evaluation metric, measuring the degree of fluctuation in the accuracy of the five outputs. The CV is defined as follows

$$CV = \frac{\sigma}{\mu} \tag{30}$$

where $\sigma$ denotes the standard deviation of $Accuracy_{Geom}$ or $Accuracy_{Num/Text/Boolean}$, and $\mu$ represents their mean. A smaller CV indicates greater consistency across multiple generations. In addition, the Stability-Adjusted Accuracy (SAA) metric is introduced:

$$SAA = \frac{pass@5}{1 + CV} \tag{31}$$

which combines pass@5 with CV to penalize models exhibiting poor stability. A higher pass@5 coupled with a lower CV jointly contributes to a higher SAA score, enabling this metric to more comprehensively reflect overall model performance.

Robustness testing evaluates the stability and resilience of model performance when natural language descriptions vary. As defined in **Sections 2.2.2 and 2.2.3**, task descriptions are categorized into explicit prompts and underspecified prompts. Robustness evaluation focuses on the underspecified prompts, recording the accuracy of each model on such tasks and comparing the results against those obtained under explicit prompts. This comparison captures the extent of performance degradation and adaptability of models when confronted with semantic ambiguity.

### 3.2.6. Rank

This study employs the Entropy Weight Method (EWM) to determine the weights of evaluation metrics and calculate the overall model scores based on multiple performance indicators across various capability levels. EWM is an objective weighting method based on information entropy, where the entropy of each metric is calculated to measure its variability. The greater the variability, the more information it contains, thus increasing its weight and contributing more significantly to the overall evaluation. This method avoids biases introduced by subjective weighting, ensuring a more fair and scientific distribution of weights. Ultimately, a relative ranking of model capabilities is produced based on the weighted results.

$$S_i = \sum_{j=1}^{n} \frac{(1 - e_j)}{\sum_{j=1}^{n} (1 - e_j)} \cdot x_{ij} \tag{32}$$

Specifically, $S_i$ represents the overall score of the $i$-th model (or scheme), $x_{ij}$ denotes the normalized score of the $i$-th model on the $j$-th metric, $m$ is the number of evaluated objects, $n$ is the number of evaluation metrics, and $e_j$ is the information entropy of the $j$-th metric, reflecting the degree of uncertainty of the metric across all objects:

$$e_j = -\frac{1}{\ln m} \sum_{i=1}^{m} p_{ij} \ln p_{ij}, p_{ij} = \frac{x_{ij}}{\sum_{i=1}^{m} x_{ij}} \tag{33}$$

From this, the differentiation coefficient $d_j = 1 - e_j$ is derived, reflecting the degree of distinction for the metric; the weight w_j is then calculated as $w_j = \frac{d_j}{\sum_{j=1}^{n} d_j}$. Finally, the overall score S_i for each model is obtained through the weighted sum of the metrics, where metrics with higher weights exert a greater influence on the evaluation, thus enabling an objective ranking of model performance.This study first calculates the entropy-weighted scores independently for the three types of questions in GeoSQL-Bench. The scores for the three question types are then used as new evaluation metrics, and a secondary weighting integration using EWM is applied to obtain the final overall performance scores for each model and their ranking.

### 3.3. Resource Consumption and Error Type Logging

In addition to evaluating execution result accuracy, GeoSQL-Eval also records resource consumption and error types to support cross-model performance comparison and diagnostics. In terms of resource consumption, the number of tokens and generation time for models are recorded separately for multiple choice and true/false questions, as well as syntax-level SQL generation questions. By incrementally recording and averaging these metrics, the cost-effectiveness of different models in task generation is assessed. Error types are categorized by experts based on error messages, and automatic sorting is achieved through regular expression matching. This allows for the quantification of model performance across different error categories and provides a basis for future optimization. The error type classification is shown in **Table 5**.

Table 5 Error Type Classification in GeoSQL-Eval

| ID | Category | Description |
| --- | --- | --- |
| 1 | Environment/Connection Errors | Session interruptions, driver failures, or character set decoding issues preventing execution or result parsing. |
| 2 | SQL Syntax Errors | Invalid syntax, missing aliases, or illegal identifiers/literals. |
| 3 | Missing Objects | Referenced tables, views, columns, or types not found or unresolved. |
| 4 | Function/Operator Errors | Nonexistent functions, overload resolution failures, or argument type/count mismatches. |
| 5 | Geometry Parsing Errors | WKT/EWKT/WKB/BOX3D/HEX geometry string or binary parsing failures. |
| 6 | SRID/Dimension Mismatch | Missing/mixed |

### 3.4. Evaluation Strategy

This chapter conducts an evaluation of the latest models as of August 2025, based on the GeoSQL-Bench benchmark and the GeoSQL-Eval framework, and systematically explains the evaluation strategies employed. The content includes key information such as the models under evaluation, hardware configurations used for testing, and execution time, with the aim of lowering the barrier for experimental replication.

### 3.5. Models to be Evaluated

This study evaluates the latest six categories of representative models as of August 2025, including: 5 general non-reasoning models, 7 general reasoning-enhanced models, 3 general code generation models, 1 geospatial code generation model, 6 general SQL generation models, and 2 publicly available strategies for generating GeoSQL.

The model selection criterion is to prioritize the latest available versions with the largest parameter size that can be locally deployed in the research team's environment, within the same product generation from the same development team. Considering that model performance is generally positively correlated with parameter scale, all models except for General SQL Generation models (which cover different versions with varying parameter sizes) were chosen based on their largest parameter versions to assess their performance under optimal configurations. It should be emphasized that some models, although released in 2023 or 2024, are not considered outdated technologically. This is because the subsequent versions in the series have not yet been released, making these models the latest representatives of their respective categories. Meanwhile, there are currently no end-to-end vertical domain models specifically designed for GeoSQL. Therefore, this study only evaluates two representative strategy frameworks and strictly follows the experimental parameters set forth in the original literature for configuration and testing. The selected strategies and models are listed in **Table 6**.

**Table 6 Information of evaluated LLMs and strategies.** "N/A" indicates that the parameter size of the model was not publicly released at the time of publication and is therefore marked as unknown.

| Category | Name | Developer | Parameter/Notes | Data |
|---|---|---|---|---|
| General Non-Reasoning | **Claude3.7-Sonnet | Anthropic | N/A | 202502 |
|  | **DeepSeek-V3-0324 | DeepSeek | 685B | 202503 |
|  | **GPT-4.1 | OpenAI | N/A | 202504 |
|  | **GPT-4.1-mini | OpenAI | N/A | 202504 |
|  | Qwen-3 | Alibaba | 32B | 202504 |
| General Reasoning-Enhanced | **DeepSeek-R1-0528 | DeepSeek | 671B | 202505 |
|  | **Gemini2.5-Flash-0520 | Google | N/A | 202505 |
|  | **GPT-5 | OpenAI | N/A | 202508 |
|  | GPT-OSS | OpenAI | 20B | 202508 |
|  | **o4-mini | OpenAI | N/A | 202504 |
|  | Qwen-3-Thinking | Alibaba | 32B | 202507 |
|  | **QwQ-32B | Alibaba | 32B | 202503 |
| General Code Generation | Code-Llama | Meta | 13B | 202308 |
|  | DeepSeek-Coder-V2 | DeepSeek | 16B | 202406 |
|  | **Qwen2.5-Coder | Alibaba | 32B | 202411 |
| Geospatial Code Generation | GeoCode-GPT-7B | WHU | 7B | 202503 |
| General SQL Generation | CodeS[44] | RUC | 3B/7B/15B | 202410 |
|  | XiYan-SQL[45] | Alibaba | 7B/14B/32B | 202411 |
| GeoSQL Generation Strategies | **Monkuu[33] | CUG | Strategy | 202507 |
|  | **SpatialSQL[21] | TJNU | Strategy | 202508 |

### 3.6. Setup Configuration

This study divides the inference methods into two categories. The first category consists of commercial closed-source models, which are tested via their official API interfaces and charged based on the number of API calls. In Table 6, these models are marked with double asterisks (**), and the specific billing standards can be found in the latest instructions on each model's official website (these standards are subject to dynamic adjustment over time). It should be noted that Monkuu and SpatialSQL are not independently built LLMs. Instead, they are implemented based on base models through RAG or prompt strategies, with the base models relying on the commercial closed-source GPT-4o-mini and GPT-4 models. Therefore, these are also categorized under the API call models in this study. The second category is local inference deployment, with an experimental environment configured with 8 NVIDIA A800 GPUs, each having 80 GB of memory, and a 64-core CPU, ensuring stability and efficiency during the inference process. To ensure comparability between different models, the temperature parameter is uniformly set to 0.2. This relatively low value enhances the determinism of the generated results while retaining some randomness to simulate the uncertainty of user queries in real-world scenarios. Additionally, considering the differences in context handling capabilities across model types, the maximum context length for non-reasoning models is set to 1024 tokens, while the maximum context length for reasoning models is set to 8096 tokens, in order to fully utilize the performance of each model type.

### 3.7. Time Consumption

The experimental process of this study spans the entire cycle from data source collection to result analysis, including data source gathering, test set creation, expert verification of the test set, model evaluation and result statistics, and result analysis. The time distribution for each stage is shown in **Table 7**.

**Table 7** Time Distribution Across Experimental Stages

| Stage | Time (hours) |
| --- | --- |
| Data collection | 96 |
| Test set construction | 640 |
| Expert validation of test set | 224 |
| Model evaluation and results aggregation | 168 |
| Results analysis | 48 |
| **Total** | **1176** |

### 4. Result and Analysis

This study conducts an evaluation centered on five major capability levels, and calculates and analyzes the entropy-weighted comprehensive scores, overall scores, and rankings for each question type based on the three categories in the GeoSQL-Bench. Finally, a comparative analysis of resource consumption and error types across different models is provided.

### 4.1. Conceptual Understanding Ability

Based on the overall accuracy and average performance of Multiple Choice and True/False questions (see **Table 9**), general reasoning-enhanced models and general non-reasoning models demonstrate a clear overall advantage, with most achieving an average accuracy above 0.85. In contrast, general code generation models and geospatial code generation models lag behind, with rule-following accuracy generally below 0.70, indicating their limited transferability in GeoSQL cognitive tasks. General SQL generation models exhibit a notable scale effect: large-scale models such as XiYan-SQL-32B achieve an average accuracy of 0.855, while smaller models like CodeS-7B only reach 0.178, almost failing entirely on certain question types. The GeoSQL strategy framework, relying on general non-reasoning models, shows relatively stable overall performance. Rule-following (ACC_RULE) remains a common weakness, with even the best-performing model, DeepSeek-R1-0528, reaching only 0.809, while most models fall within the 0.70–0.78 range. This result reflects a deficiency in models' understanding of PostGIS function specifications and details, which serves as a key bottleneck in further improving GeoSQL cognitive abilities.

**Table 9 presents the evaluation results for Multiple Choice and True/False questions.** ACC_FUNC denotes Function Purpose Recognition, ACC_PARAM denotes Parameter Matching Recognition, ACC_TYPE denotes

Return Type Recognition, ACC_RULE denotes General Rule Compliance, and AVG represents the average score across these four categories.

| Category | Name | ACC_FUNC | ACC_PARAM | ACC_TYPE | ACC_RULE | AVG |
|---|---|---|---|---|---|---|
| 1 | Claude3.7-Sonnet | 0.959 | 0.879 | 0.974 | 0.782 | 0.899 |
| 1 | DeepSeek-V3-0324 | 0.957 | 0.832 | 0.974 | 0.715 | 0.869 |
| 1 | GPT-4.1 | 0.938 | 0.894 | 0.953 | 0.759 | 0.886 |
| 1 | GPT-4.1-mini | 0.924 | 0.818 | 0.944 | 0.746 | 0.858 |
| 1 | Qwen3-32B | 0.931 | 0.878 | 0.950 | 0.757 | 0.879 |
| 2 | DeepSeek-R1-0528 | 0.947 | 0.901 | 0.994 | 0.809 | 0.913 |
| 2 | Gemini2.5-Flash-0520 | 0.921 | 0.828 | 0.973 | 0.770 | 0.873 |
| 2 | GPT-5 | 0.969 | 0.882 | 0.950 | 0.774 | 0.894 |
| 2 | GPT-OSS-20B | 0.901 | 0.810 | 0.964 | 0.693 | 0.842 |
| 2 | o4-mini | 0.896 | 0.824 | 0.952 | 0.776 | 0.862 |
| 2 | Qwen3-32B-Thinking | 0.916 | 0.836 | 0.973 | 0.745 | 0.867 |
| 2 | QwQ-32B | 0.922 | 0.772 | 0.950 | 0.735 | 0.845 |
| 3 | Code-Llama-13B | 0.778 | 0.487 | 0.771 | 0.522 | 0.639 |
| 3 | DeepSeek-Coder-V2-16B | 0.615 | 0.611 | 0.822 | 0.679 | 0.682 |
| 3 | Qwen2.5-Coder-32B | 0.900 | 0.821 | 0.953 | 0.712 | 0.846 |
| 4 | GeoCode-GPT-7B | 0.797 | 0.651 | 0.726 | 0.599 | 0.693 |
| 5 | CodeS-15B | 0.626 | 0.278 | 0.159 | 0.286 | 0.337 |
| 5 | CodeS-3B | 0.046 | 0.532 | 0.204 | 0.510 | 0.323 |
| 5 | CodeS-7B | 0.474 | 0.103 | 0.135 | 0.000 | 0.178 |
| 5 | XiYan-SQL-14B | 0.896 | 0.788 | 0.932 | 0.713 | 0.832 |
| 5 | XiYan-SQL-32B | 0.907 | 0.834 | 0.953 | 0.726 | 0.855 |
| 5 | XiYan-SQL-7B | 0.876 | 0.751 | 0.903 | 0.657 | 0.797 |
| 6 | Monkuu | 0.890 | 0.722 | 0.906 | 0.690 | 0.802 |
| 6 | SpatialSQL | 0.909 | 0.821 | 0.950 | 0.725 | 0.851 |

**4.2. Structured SQL Generation Ability**

As shown in **Table 10**, there are significant differences between models in terms of execution pass rate (EPR) and syntax accuracy (SA). General non-reasoning models are generally robust, with SA values typically close to 0.99. However, execution performance varies significantly, with Claude3.7-Sonnet leading at 0.877, while Qwen3-32B ranks last at only 0.763. Reasoning-enhanced models are generally stronger, with o4-mini and DeepSeek-R1-0528 consistently at the top, the latter achieving the highest score in Table Schema Retrieval (0.919). Overall, syntax accuracy does not guarantee execution within the database; execution reliability remains a key bottleneck in GeoSQL tasks.

**Table 10. Execution Pass Rate(EPR) and Syntax Accuracy Results(SA).** S_EPR and S_SA correspond to Execution Pass Rate and Syntax Accuracy for Syntax-level SQL Generation, while T_EPR and T_SA correspond to the same metrics for Table Schema Retrieval.

| Category | Name | S_EPR | S_SA | T_EPR | T_SA |
|---|---|---|---|---|---|
| 1 | Claude3.7-Sonnet | 0.877 | 0.990 | 0.772 | 0.987 |
| 1 | DeepSeek-V3-0324 | 0.873 | 0.994 | 0.715 | 0.988 |
| 1 | Gemini2.5-Flash-0520 | 0.827 | 0.986 | 0.680 | 0.973 |

| Category | Name | S_EPR | S_SA | T_EPR | T_SA |
| --- | --- | --- | --- | --- | --- |
| 1 | GPT-4.1 | 0.846 | 0.982 | 0.773 | 0.980 |
| 1 | GPT-4.1-mini | 0.780 | 0.977 | 0.672 | 0.988 |
| 1 | Qwen3-32B | 0.763 | 0.982 | 0.774 | 0.969 |
| 2 | DeepSeek-R1-0528 | 0.812 | 0.993 | 0.919 | 0.977 |
| 2 | Gemini2.5-Flash-0520 | 0.827 | 0.986 | 0.680 | 0.973 |
| 2 | GPT-5 | 0.856 | 0.976 | 0.711 | 0.988 |
| 2 | GPT-OSS-20B | 0.689 | 0.979 | 0.814 | 0.978 |
| 2 | o4-mini | 0.821 | 0.995 | 0.875 | 0.993 |
| 2 | Qwen3-32B-Thinking | 0.747 | 0.934 | 0.649 | 0.779 |
| 2 | QwQ-32B | 0.718 | 0.984 | 0.731 | 0.919 |
| 3 | Code-Llama-13B | 0.486 | 0.783 | 0.547 | 0.989 |
| 3 | DeepSeek-Coder-V2-16B | 0.668 | 0.933 | 0.744 | 0.986 |
| 3 | Qwen2.5-Coder-32B | 0.731 | 0.925 | 0.573 | 0.937 |
| 4 | GeoCode-GPT-7B | 0.363 | 0.826 | 0.409 | 0.923 |
| 5 | CodeS-15B | 0.622 | 0.966 | 0.098 | 0.183 |
| 5 | CodeS-3B | 0.382 | 0.866 | 0.365 | 0.734 |
| 5 | CodeS-7B | 0.581 | 0.942 | 0.162 | 0.297 |
| 5 | XiYan-SQL-14B | 0.461 | 0.800 | 0.442 | 0.792 |
| 5 | XiYan-SQL-32B | 0.773 | 0.962 | 0.834 | 0.986 |
| 5 | XiYan-SQL-7B | 0.520 | 0.855 | 0.620 | 0.893 |
| 6 | Monkuu | 0.652 | 0.926 | 0.779 | 0.977 |
| 6 | SpatialSQL | 0.847 | 0.987 | 0.863 | 0.994 |

**4.3. Semantic Alignment and Invocation Ability**

**Figures 11 and 12** present scatter plot comparisons of the performance of different models in terms of semantic accuracy for syntax-level SQL generation questions and table schema retrieval questions, respectively. In the syntax-level tasks, the models are generally distributed evenly, but the mean performance is relatively low: the function hit rate is slightly above 0.5, while the argument match accuracy is only around 0.25. General non-reasoning models generally perform better, with most distributed in the upper-right region. Notably, Claude3.7-Sonnet and DeepSeek-V3-0324 rank at the top. Reasoning-enhanced models show clear variation, with GPT-5 performing on par with top general models, while QwQ-32B and GPT-OSS-20B lag significantly behind. General code generation models and smaller-scale SQL models are concentrated in the lower-left region, exhibiting weaker overall performance. The large-scale SQL model XiYan-SQL-32B and the GeoSQL strategy model SpatialSQL are positioned above the mean line, reflecting the advantages of model scale and strategic mechanisms in function recognition tasks. In the table schema level tasks, the distribution is more concentrated, with most models clustering above the mean line, indicating that table structure and field recognition capabilities are relatively mature. The main differences arise from function hit rates. GPT-5 and o4-mini perform the best, while a few smaller general SQL models and code generation models lower the overall distribution, particularly in function recognition tasks.

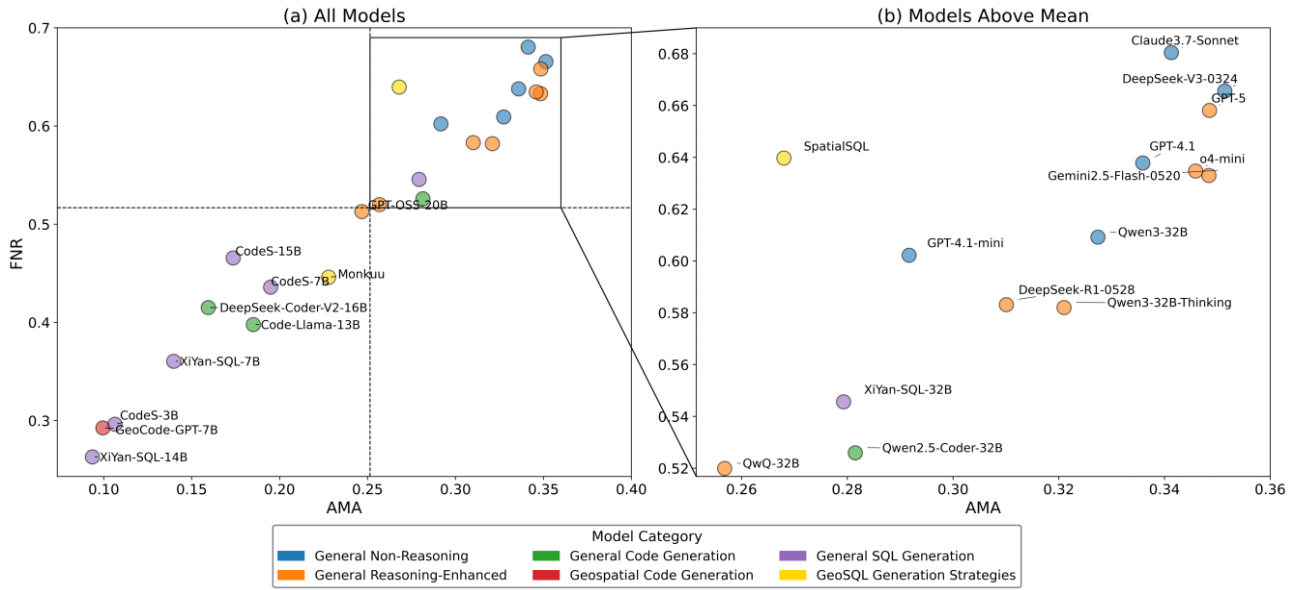

**Figure 11. Function Name Rate (FNR) and Argument Match Accuracy (AMA) results for syntax-level SQL generation questions.** Panel (a) presents the results for all models, while panel (b) highlights models with above-mean performance, with the corresponding range marked by the rectangle in panel (a). The circles are of uniform size and do not represent any additional metric.

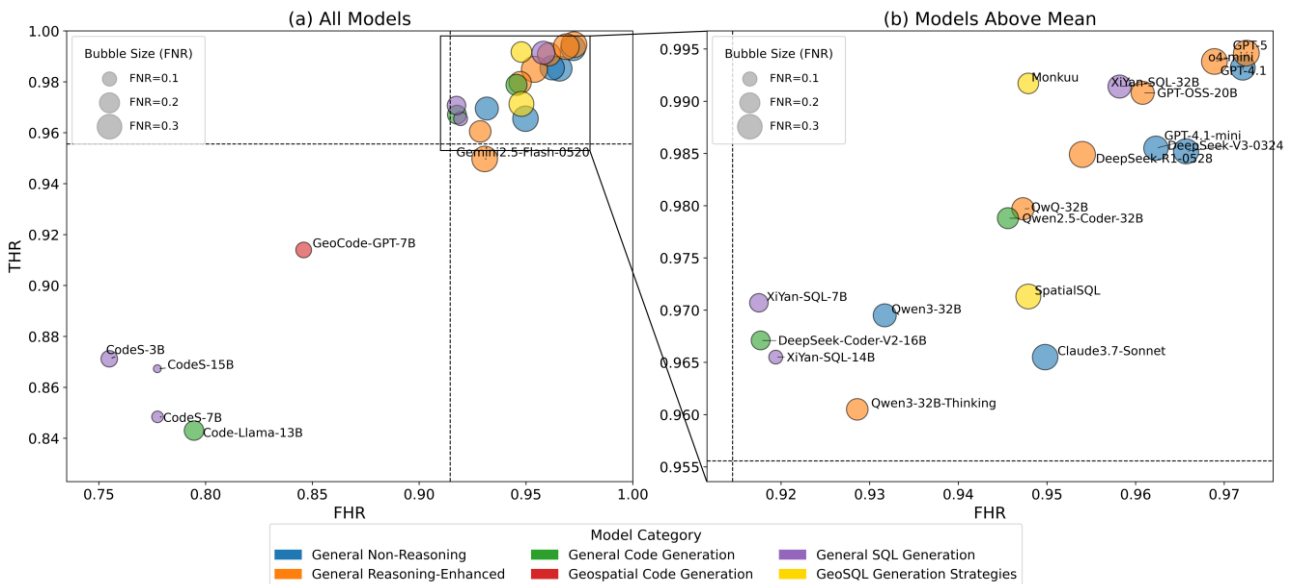

**Figure 12. Table Hit Rate (THR), Field Hit Rate (FHR), and Function Name Rate (FNR) results for table schema retrieval questions.** Panel (a) shows the results for all models, while panel (b) highlights models with above-mean performance, with the corresponding range indicated by the rectangle in panel (a). In both panels, the size of each scatter point represents the Function Name Rate (FNR).

### 4.4. Execution and Answer Accuracy Ability

In the syntax-level SQL generation question task (see **Figure 13**), general non-reasoning models lead overall, with Claude3.7-Sonnet at the top. Among reasoning-enhanced models, o4-mini and GPT-5 stand out with exceptional performance. Geometric accuracy is generally above 0.70, while accuracy for numeric and textual types is lower, which results in an overall decrease in ACC_ALL. In the table schema retrieval question task (see **Figure 14**), the overall accuracy is significantly lower than in the previous task. Reasoning-enhanced models lead in ACC_ALL, with o4-mini achieving 0.4612, ranking first, and DeepSeek-R1-0528 surpassing 0.44. Although GPT-

5's ACC_ALL is only 0.3806, its ACC_Geo reaches 0.9293, the highest among all models, highlighting the advantage of reasoning capabilities in geometric tasks. It is worth noting that since this task relies on row and column matching, non-geometric errors are minimal once the matching is consistent, with most models achieving near-perfect accuracy for non-geometric types. The differences are limited, and thus, no visualization is provided for these results.

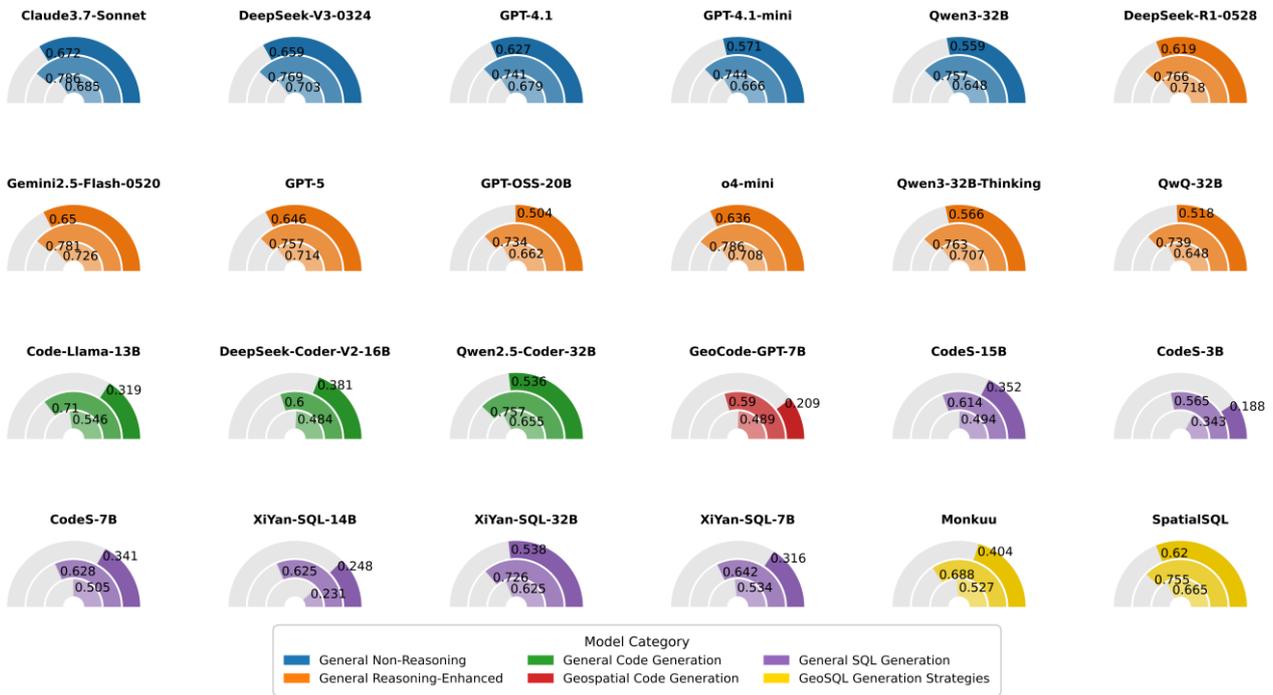

**Figure 13. Overall Accuracy (ACC_ALL), Geometric Accuracy (ACC_Geo), and Accuracy on Numeric, Text, and Boolean Types (ACC_Other) results for syntax-level SQL generation questions.** From the outer to the inner rings, the first ring represents overall accuracy, the second ring represents geometric accuracy, and the third ring represents the accuracy of numeric, text, and boolean types.

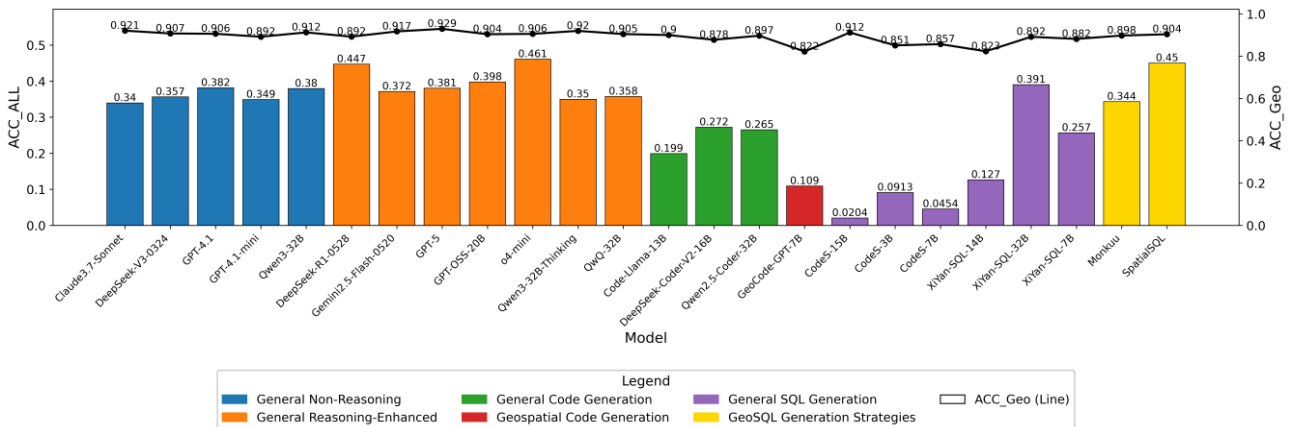

**Figure 14. Overall Accuracy (ACC_ALL), Geometric Accuracy (ACC_Geo) results for table schema retrieval questions.** The bars represent ACC_ALL, while the line represents ACC_Geo.

**4.5. Generalization and Robust Reasoning Ability**

**Table 11** and **Table 12** display the performance of different models in two tasks, showing pass@n, CV, and SAA.

**Table 11. Results of syntax-level SQL generation questions.** The values in parentheses under pass@3 indicate the improvement over pass@1, and those under pass@5 indicate the improvement over pass@3.

| Category | Name | pass@1 (%) | pass@3 (%) | pass@5 (%) | CV | SAA |
|---|---|---|---|---|---|---|
| 1 | Claude3.7-Sonnet | 67.03 | 70.83(+3.80) | 71.98(+1.15) | 0.0302 | 69.87 |

| Category | Name | pass@1 (%) | pass@3 (%) | pass@5 (%) | CV | SAA |
|---|---|---|---|---|---|---|
| 1 | DeepSeek-V3-0324 | 66.15 | 70.83(+4.68) | 72.36(+1.53) | 0.0379 | 69.72 |
| 1 | GPT-4.1 | 63.02 | 66.20(+3.18) | 67.22(+1.02) | 0.0273 | 65.43 |
| 1 | GPT-4.1-mini | 57.61 | 61.52(+3.91) | 63.23(+1.71) | 0.0387 | 60.87 |
| 1 | Qwen3-32B | 55.93 | 58.95(+3.02) | 60.05(+1.10) | 0.0299 | 58.31 |
| 2 | DeepSeek-R1-0528 | 61.63 | 70.08(+8.45) | 72.44(+2.36) | 0.0682 | 67.82 |
| 2 | Gemini2.5-Flash-0520 | 65.88 | 85.31(+19.43) | 89.86(+4.55) | 0.1294 | 79.56 |
| 2 | GPT-5 | 65.99 | 83.03(+17.04) | 88.04(+5.01) | 0.1194 | 78.65 |
| 2 | GPT-OSS-20B | 50.04 | 59.89(+9.85) | 63.58(+3.69) | 0.0988 | 57.86 |
| 2 | o4-mini | 63.77 | 84.59(+20.82) | 89.8(+5.21) | 0.1417 | 78.65 |
| 2 | Qwen3-32B-Thinking | 55.74 | 69.23(+13.49) | 73.03(+3.80) | 0.1124 | 65.65 |
| 2 | QwQ-32B | 51.70 | 63.50(+11.8) | 66.93(+3.4) | 0.1074 | 60.44 |
| 3 | Code-Llama-13B | 32.11 | 45.36(+13.25) | 51.40(+6.04) | 0.1875 | 43.28 |
| 3 | DeepSeek-Coder-V2-16B | 38.19 | 41.66(+3.47) | 43.00(+1.34) | 0.0495 | 40.97 |
| 3 | Qwen2.5-Coder-32B | 53.63 | 57.24(+3.61) | 58.58(+1.34) | 0.037 | 56.49 |
| 4 | GeoCode-GPT-7B | 21.09 | 26.63(+5.54) | 29.11(+2.48) | 0.1309 | 25.74 |
| 5 | CodeS-15B | 34.87 | 43.11(+8.244) | 46.27(+3.16) | 0.116 | 41.46 |
| 5 | CodeS-3B | 18.14 | 28.23(+10.09) | 32.67(+4.44) | 0.2308 | 26.54 |
| 5 | CodeS-7B | 33.77 | 42.23(+8.46) | 44.96(+2.73) | 0.1181 | 40.21 |
| 5 | XiYan-SQL-14B | 24.99 | 30.51(+5.52) | 32.75(+2.24) | 0.1109 | 29.48 |
| 5 | XiYan-SQL-32B | 53.97 | 56.44(+2.47) | 57.56(+1.12) | 0.0268 | 56.06 |
| 5 | XiYan-SQL-7B | 31.50 | 35.51(+4.01) | 37.57(+2.06) | 0.0723 | 35.04 |
| 6 | Monkuu | 40.86 | 62.40(+21.54) | 70.83(+8.43) | 0.2175 | 58.18 |
| 6 | SpatialSQL | 61.65 | 77.07(+15.42) | 82.15(+5.08) | 0.1184 | 73.45 |

**Table 11. Results of table schema retrieval questions.** The values in parentheses under pass@3 indicate the improvement over pass@1, and those under pass@5 indicate the improvement over pass@3.

| Category | Name | pass@1 (%) | pass@3 (%) | pass@5 (%) | CV | SAA |
|---|---|---|---|---|---|---|
| 1 | Claude3.7-Sonnet | 33.78 | 36.06(+2.28) | 36.47(+0.41) | 0.0334 | 35.29 |
| 1 | DeepSeek-V3-0324 | 35.45 | 40.46(+5.01) | 42.04(+1.58) | 0.0715 | 39.23 |
| 1 | GPT-4.1 | 37.87 | 44.32(+6.45) | 46.45(+2.13) | 0.0851 | 42.81 |
| 1 | GPT-4.1-mini | 34.76 | 37.73(+2.97) | 38.79(+1.06) | 0.046 | 37.08 |
| 1 | Qwen3-32B | 38.42 | 41.48(+3.06) | 42.83(+1.35) | 0.0451 | 40.98 |
| 2 | DeepSeek-R1-0528 | 44.41 | 54.15(+9.74) | 57.31(+3.16) | 0.1057 | 51.83 |
| 2 | Gemini2.5-Flash-0520 | 37.08 | 44.92(+7.84) | 47.94(+3.02) | 0.1057 | 43.36 |
| 2 | GPT-5 | 38.52 | 43.57(+5.05) | 45.15(+1.58) | 0.0667 | 42.33 |
| 2 | GPT-OSS-20B | 40.84 | 49.23(+8.39) | 52.44(+3.21) | 0.103 | 47.54 |
| 2 | o4-mini | 46.64 | 56.29(+9.65) | 59.91(+3.62) | 0.1032 | 54.31 |
| 2 | Qwen3-32B-Thinking | 34.85 | 49.14(+14.29) | 53.41(+4.27) | 0.1733 | 45.52 |
| 2 | QwQ-32B | 35.59 | 47.15(+11.56) | 51.14(+3.99) | 0.1478 | 44.55 |
| 3 | Code-Llama-13B | 19.81 | 24.59(+4.78) | 26.91(+2.32) | 0.1244 | 23.93 |
| 3 | DeepSeek-Coder-V2-16B | 26.91 | 29.84(+2.93) | 30.95(+1.11) | 0.0583 | 29.25 |
| 3 | Qwen2.5-Coder-32B | 26.31 | 32.34(+6.03) | 35.78(+3.44) | 0.1243 | 31.82 |
| 4 | GeoCode-GPT-7B | 10.49 | 16.47(+5.98) | 18.7(+2.23) | 0.2277 | 15.23 |
| 5 | CodeS-15B | 1.81 | 5.24(+3.43) | 7.66(+2.42) | 0.4895 | 5.14 |
| 5 | CodeS-3B | 8.91 | 15.59(+6.68) | 18.98(+3.39) | 0.2887 | 14.73 |
| 5 | CodeS-7B | 4.59 | 9.23(+4.64) | 12.71(+3.48) | 0.3761 | 9.24 |

| Category | Name | pass@1 (%) | pass@3 (%) | pass@5 (%) | CV | SAA |
|---|---|---|---|---|---|---|
| 5 | XiYan-SQL-14B | 12.95 | 15.03(+2.08) | 16.19(+1.16) | 0.091 | 14.84 |
| 5 | XiYan-SQL-32B | 39.03 | 41.16(+2.13) | 42.04(+0.88) | 0.031 | 40.78 |
| 5 | XiYan-SQL-7B | 25.89 | 29.14(+3.25) | 30.63(+1.49) | 0.0693 | 28.64 |
| 6 | Monkuu | 34.45 | 38.49(+4.04) | 42.39(+3.90) | 0.0843 | 39.09 |
| 6 | SpatialSQL | 44.92 | 49.98(+5.06) | 51.69(+1.71) | 0.0588 | 48.82 |

**Figure 15** shows the performance of different types of models in terms of pass rate. General non-reasoning models achieve relatively high results in pass@1, and the improvements in pass@3 and pass@5 are limited, as their generated results are already close to the correct answer in the first round. Reasoning-enhanced models perform best overall, with the largest improvements observed in pass@3 and pass@5. Notably, o4-mini achieves nearly 90% in pass@5 for syntax level tasks, indicating that reasoning optimization significantly enhances the model's generation diversity and problem-solving success rate. The mean line for pass@5 also reveals that general reasoning-enhanced models and GeoSQL generation strategies demonstrate significant advantages in both task types. All models perform worse on the table schema level compared to the syntax level, reflecting that complex table structure retrieval remains a major challenge for current models. The scatter plot in **Figure 16** further shows that lower CV values, along with higher SA and question accuracy, typically correspond to better performance, with data points clearly clustering in the upper-left region.

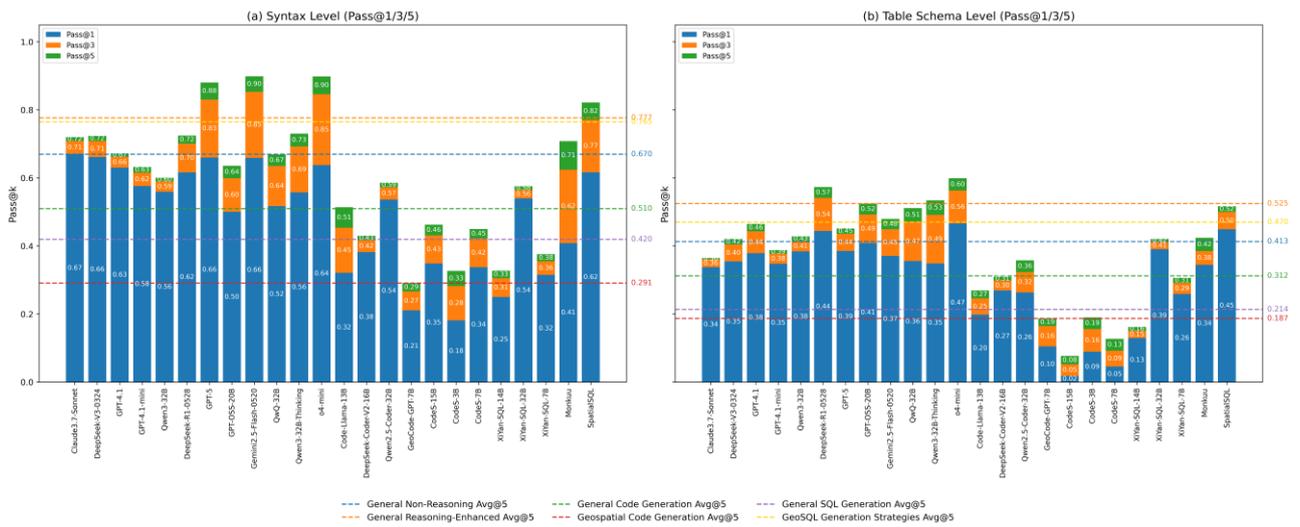

**Figure 15. Stacked bar chart of pass@n metrics.** Blue indicates the pass@1 value, orange shows the improvement of pass@3 over pass@1, and green shows the improvement of pass@5 over pass@3. The text labels within the bars display the absolute scores for pass@1, pass@3, and pass@5, respectively. Mean performance lines for different model categories are also plotted in the figure using their corresponding colors. Panel (a) presents the results for syntax-level SQL generation questions, while panel (b) presents the results for table schema retrieval questions.

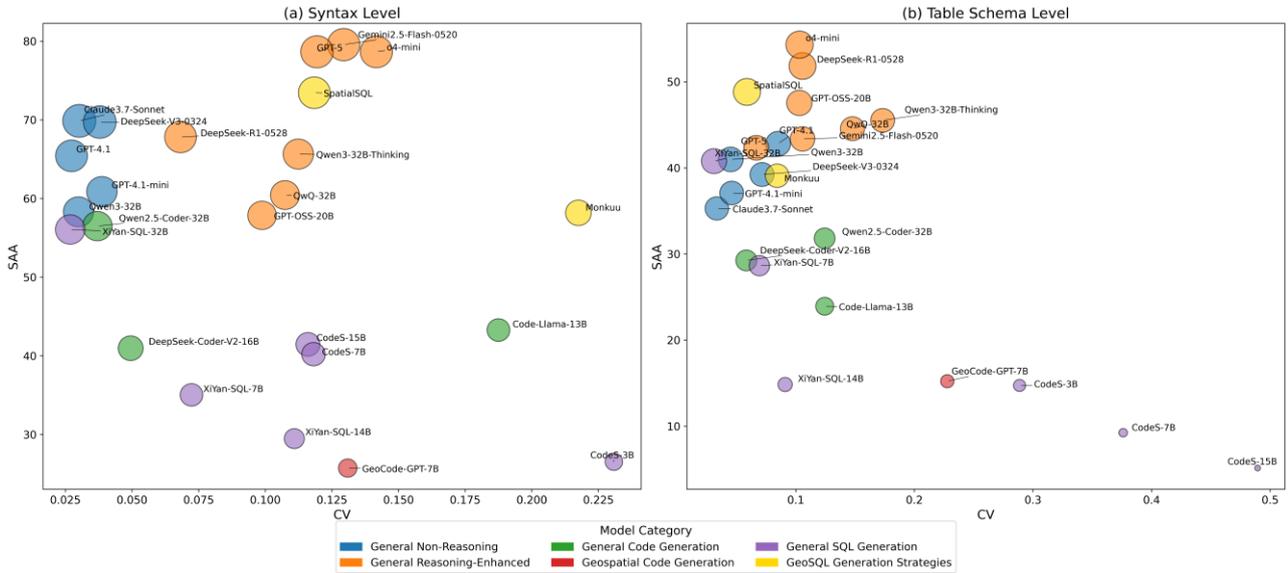

**Figure 16. Visualization of the distribution of CV and SAA.** Panel (a) shows the results for syntax-level SQL generation questions, while Panel (b) shows the results for table schema retrieval questions. In both panels, the size of each scatter point represents the accuracy on explicit questions (ALL_EXPL).

The dumbbell plot in **Figure 17** compares the accuracy differences between different categories of models for explicit and underspecified questions. The results show that underspecified questions have a more significant impact on syntax level tasks, while their effect on table schema tasks is relatively smaller. This is related to the lower baseline accuracy of the models on explicit questions. Both general non-reasoning models and general reasoning-enhanced models experience a noticeable decline when facing underspecified questions. For example, the accuracy of o4-mini decreases by 0.354 and 0.247 for explicit questions in the two tasks, respectively, indicating significant reductions and highlighting the vulnerability of these models in handling semantic ambiguity and vagueness.

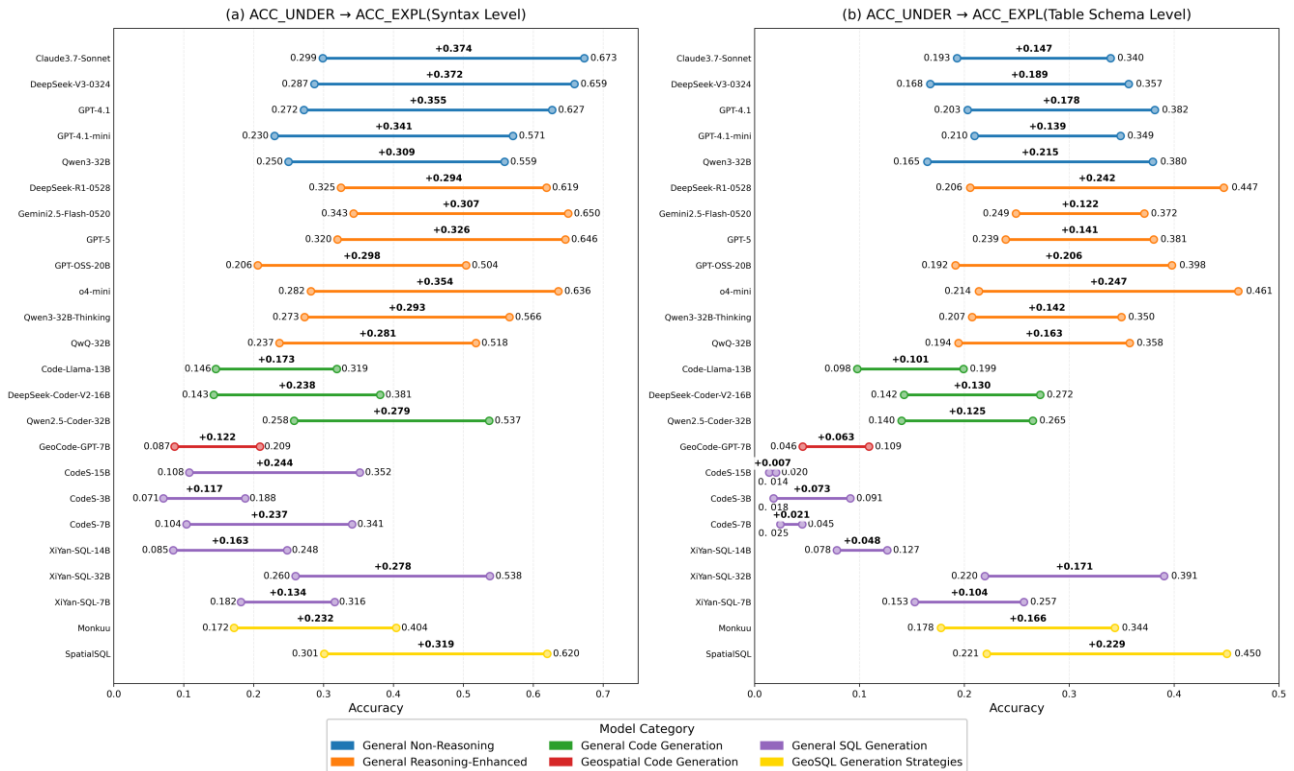

**Figure 17. Comparison of Underspecified questions accuracy (ACC_UNDER) and explicit questions accuracy (ACC_EXPL) across models.** Each horizontal bar represents the transition from ACC_UNDER (left endpoint) to ACC_EXPL (right endpoint), with the numeric labels indicating the improvement. Panel (a) shows the results for syntax-level SQL generation questions, while Panel (b) shows the results for table schema retrieval

questions. Different colors denote model categories, as indicated in the legend.

### 4.6. Rank

**Figure 18** shows the score distribution of models across four dimensions. The knowledge dimension overall favors higher score ranges, exhibiting a clear right skew, with little variation between models. The distribution in the table schema level is more concentrated compared to the syntax level, but both dimensions have large standard deviations, indicating that some models have weaknesses in these tasks while also highlighting differences between models. The standard deviation of the overall score is the largest, indicating significant differentiation in overall capabilities. Combining with the ranking heatmap in **Figure 19**, it is evident that top models perform exceptionally across multiple dimensions. For example, GPT-5, o4-mini, and DeepSeek-R1-0528 rank highly in both syntax and table schema levels, with o4-mini topping both overall and table schema level scores, highlighting its reasoning advantage in complex tasks. Overall, leading general non-reasoning models and reasoning-enhanced models exhibit significant advantages. It is worth noting that SpatialSQL ranks second in the table schema level and enters the top five in the overall ranking, demonstrating the competitive potential of GeoSQL generation strategies in specific tasks.

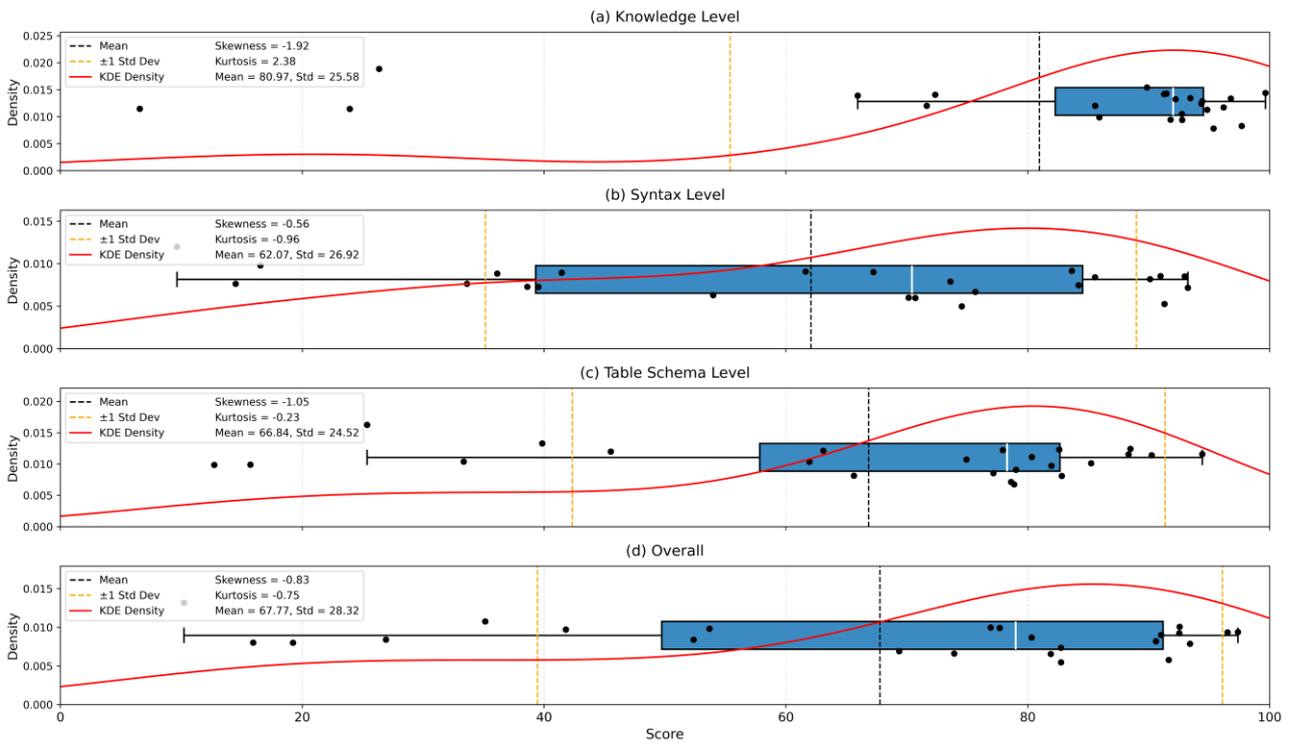

**Figure 18.** Score distributions derived from the Entropy Weight Method across different question types. Panel (a) presents the results for Multiple Choice & True/False questions, Panel (b) presents the results for syntax-level SQL generation questions, Panel (c) presents the results for table schema retrieval questions, and Panel (d) presents the overall score.

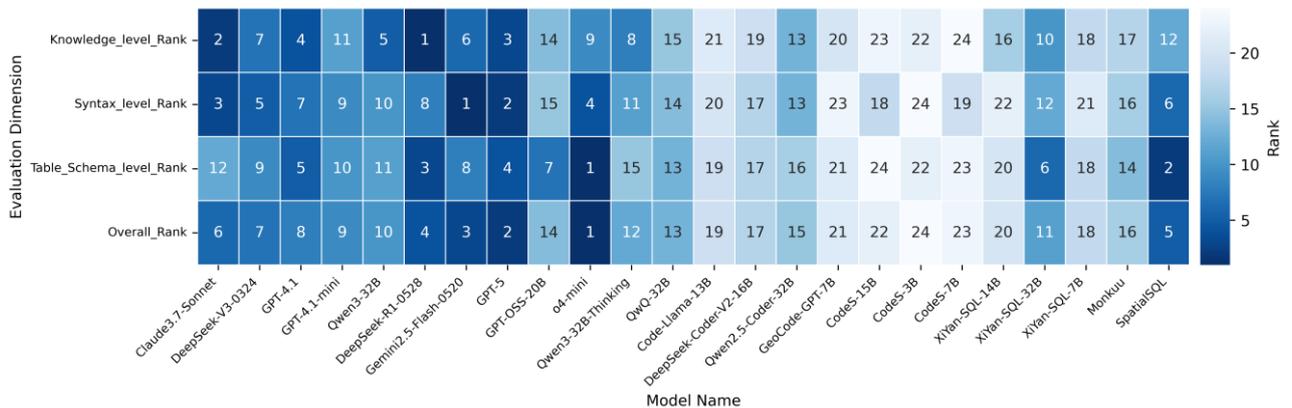

**Figure 19**. Heatmap of model rankings across different evaluation dimensions.

## 4.7. Resource & Error Logging

**Figures 20 and 21** show the distribution of different error types for each model across two tasks: syntax-level SQL generation questions and table schema retrieval questions. The figures present the proportion of each error type, highlighting the differences between models in terms of error distribution.

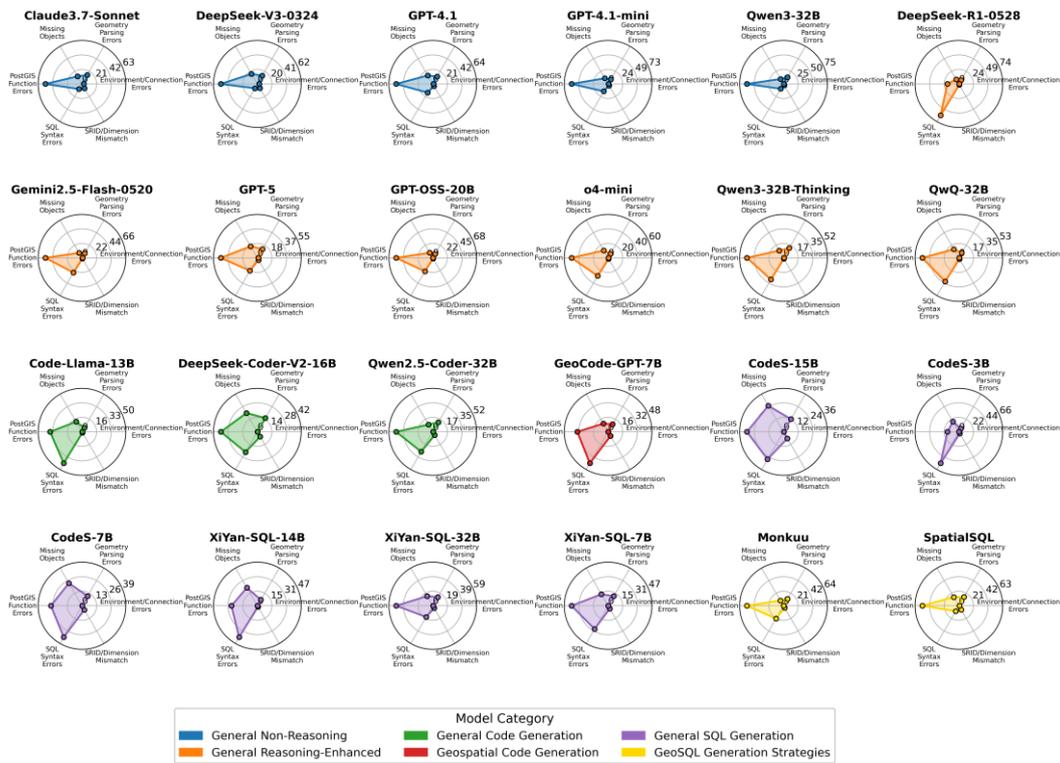

**Figure 20. Radar chart of error type distribution across different models on syntax-level SQL generation questions.** The chart depicts the relative proportions of each error type, with the values representing their percentage shares.

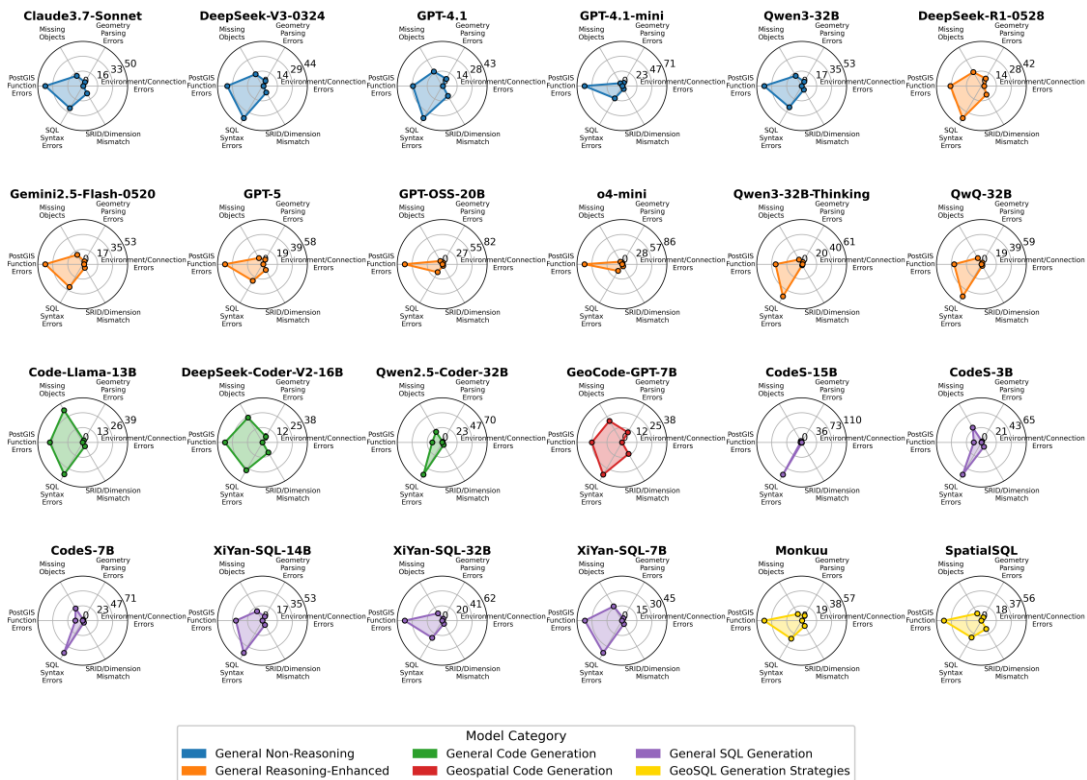

**Figure 21. Radar chart of error type distribution across different models on table schema retrieval questions.** The chart depicts the relative proportions of each error type, with the values representing their percentage shares.

As shown in **Figure 22**, although there are differences in error distribution between the syntax level and table schema level, the primary error types are consistent. Both tasks are primarily dominated by PostGIS function errors and SQL syntax errors, accounting for approximately 70% of the total. In the syntax level task, PostGIS function errors have the highest proportion, indicating that models struggle with understanding function semantics and matching parameter specifications. Although geometric parsing errors and SRID/dimension mismatches account for a low proportion, they highlight the model's weaknesses in handling spatial data and coordinate transformation. In the table schema level task, PostGIS function errors and SQL syntax errors are nearly equal in proportion. The proportion of syntax errors is about 10% higher than in the syntax level task, indicating that table structure constraints increase the complexity of syntax generation, making errors more likely in multi-table joins and complex logic mixing. At the same time, geometric parsing and SRID errors have increased compared to the syntax level. In the table schema scenario, models are more prone to issues related to spatial data formats.

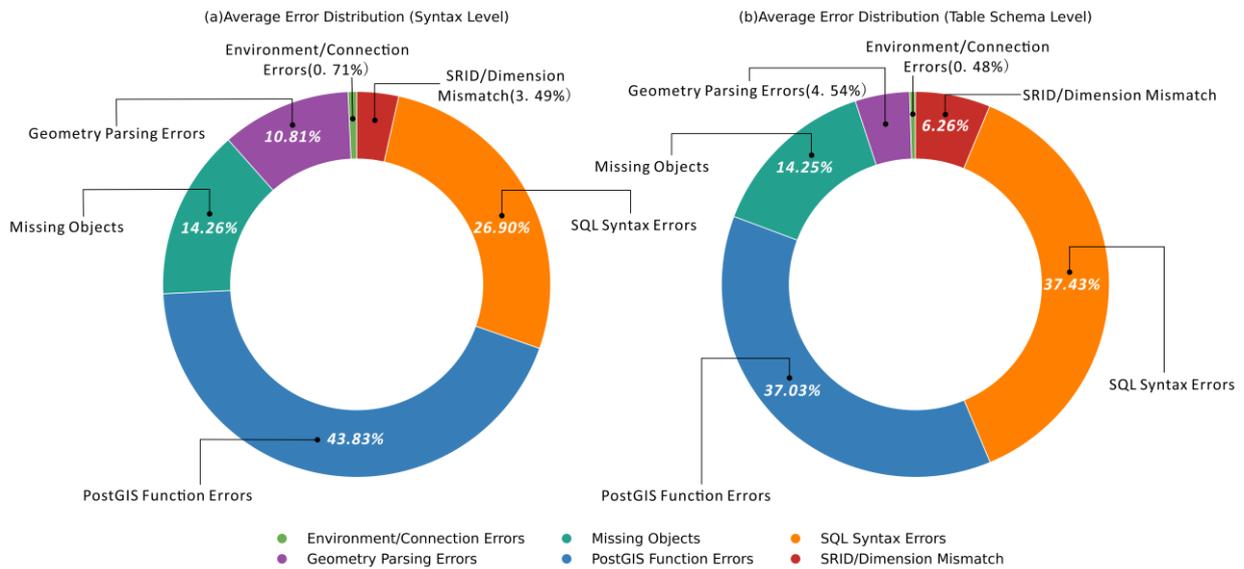

**Figure 22. Average error distributions across different error categories.** Panel (a) presents the distribution of error types for syntax-level SQL generation questions, while Panel (b) presents the distribution for table schema retrieval questions.

In terms of running efficiency, as shown in **Figures 23 and 24**, general non-reasoning models perform the best overall, while reasoning-enhanced models exhibit significant variation. Among them, o4-mini strikes a good balance between accuracy and computational overhead, delivering the best overall performance. Notably, GPT-5 can adaptively regulate the length of reasoning content based on task difficulty, and Gemini 2.5 Flash-0520, as a hybrid reasoning model, has a similar mechanism, leading to good performance in both efficiency and accuracy. In contrast, GPT-OSS-20B takes longer to process, and Qwen3-32B-Thinking consumes the most tokens, resulting in lower cost-effectiveness. Among general SQL generation models, the XiYan-SQL series significantly outperforms the CodeS series, particularly excelling in high-cost modes.

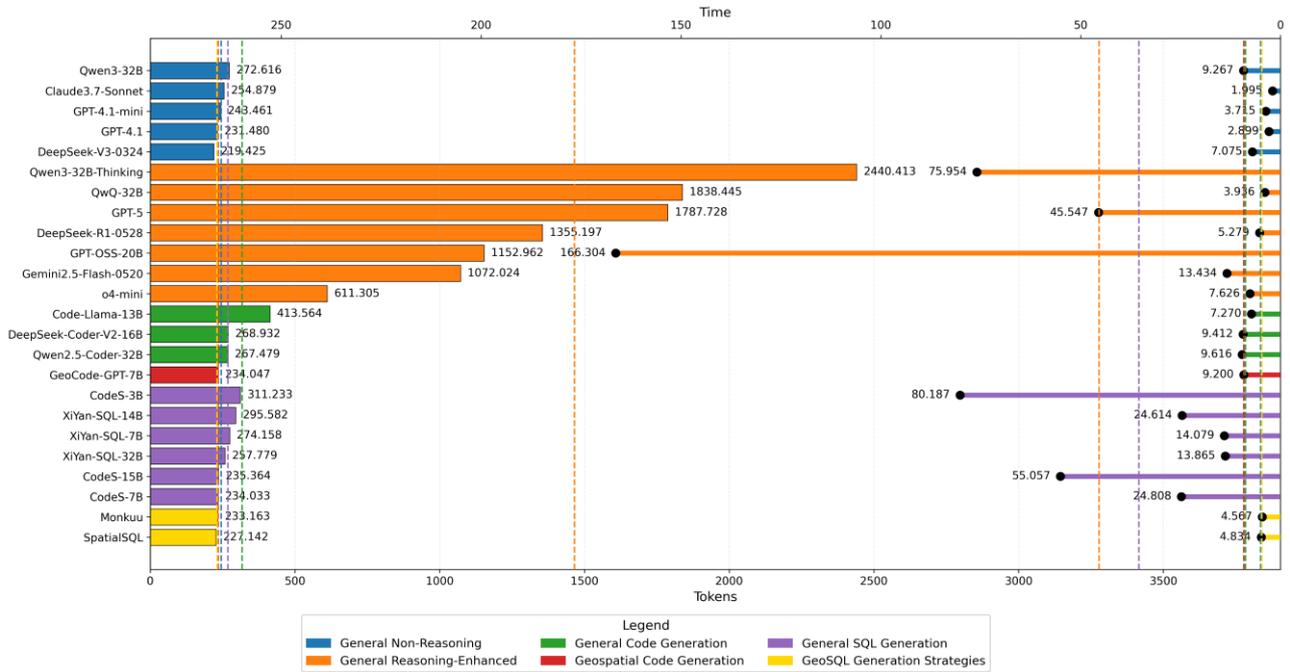

**Figure 23. Average number of tokens and average generation time for syntax-level SQL generation questions.** The bars on the left represent tokens, while the lollipops on the right represent time.

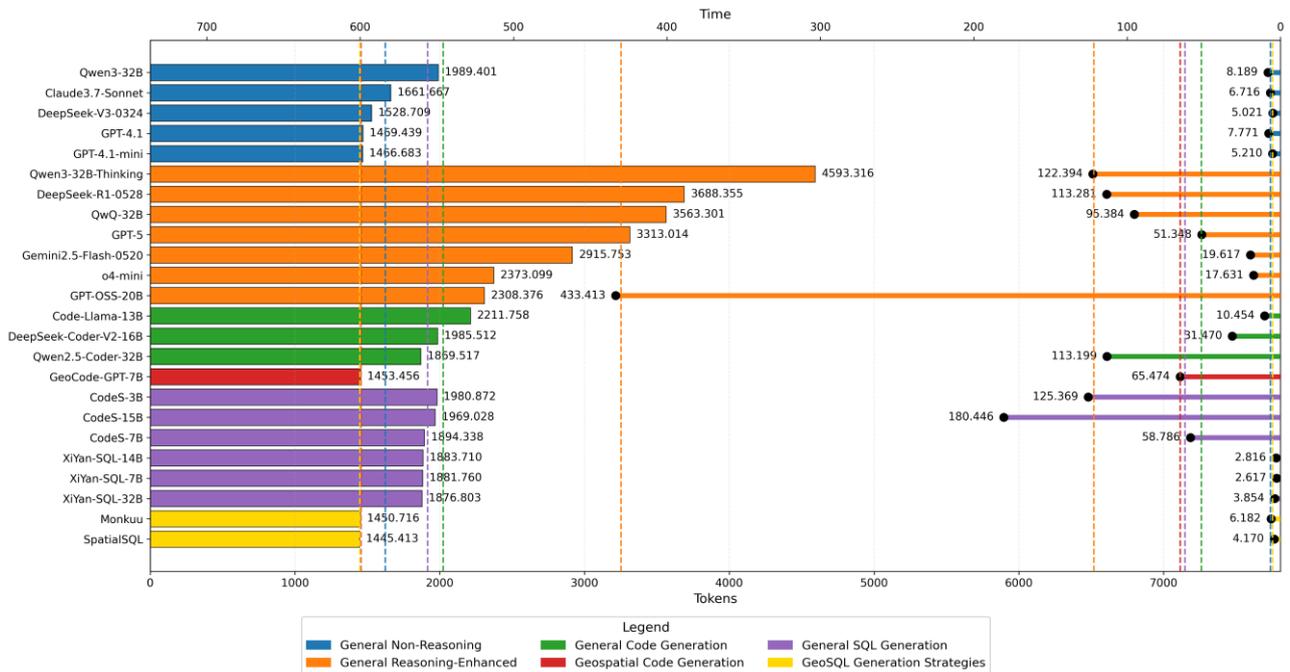

**Figure 24. Average number of tokens and average generation time for table schema retrieval questions.** The bars on the left represent tokens, while the lollipops on the right represent time.

### 4.8. GeoSQL-Eval Leaderboard Platform

We have developed and publicly released the GeoSQL-Eval leaderboard platform, as shown in **Figure 25**. This platform provides a comprehensive display of all the metrics and results for the models evaluated in this study, along with explanations for each metric and rankings. It allows research teams worldwide to submit their models for testing and supports continuous updates. The platform link is: https://haoyuejiao.github.io/GeoSQL-Eval-Leaderboard/

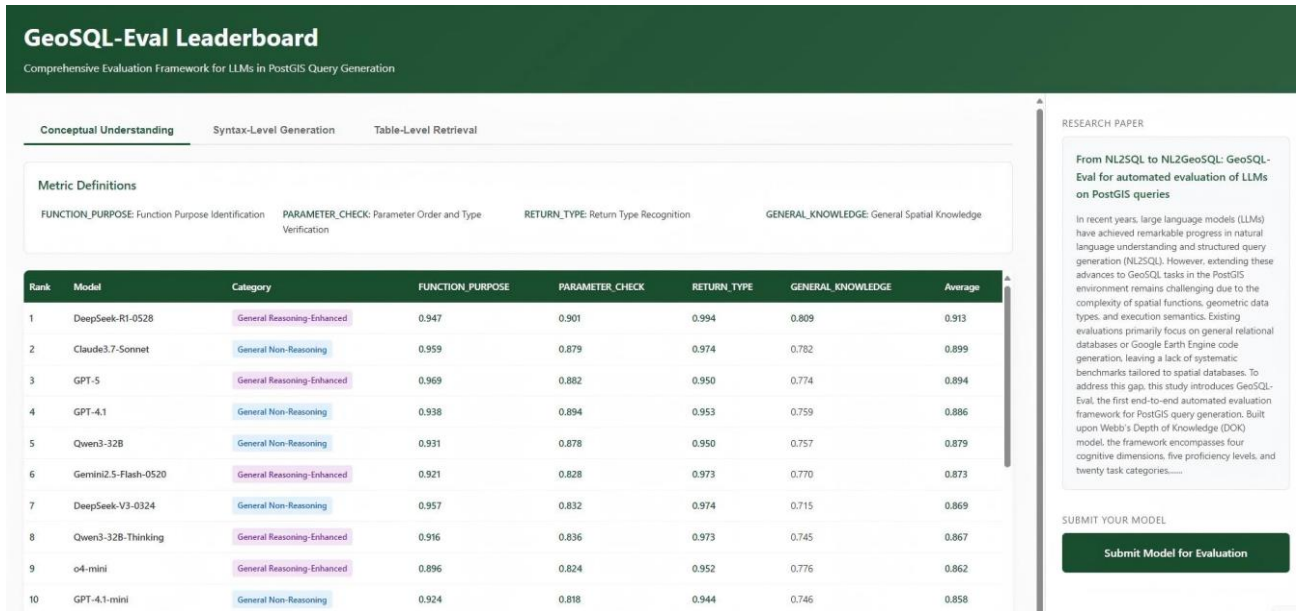

**Figure 25** The GeoSQL-Eval leaderboard platform, showcasing the evaluation results of all models, along with metric explanations and rankings.

## 5. Conclusion

This study presents GeoSQL-Eval, the first end-to-end automated evaluation framework for GeoSQL generation tasks on the PostGIS platform. The framework, implemented in Python, systematically supports multidimensional and multilayered evaluation across four cognitive dimensions, five capability levels, and 20 task categories. The framework consists of two core components: the first is the constructed benchmark set, GeoSQL-Bench, which includes 14,178 questions covering three core task types: 2,380 multiple-choice and true/false questions, 3,744 syntax generation questions, and 2,155 table schema retrieval questions. The second component is the fully automated evaluation process, which comprehensively validates the model-generated results across aspects such as knowledge mastery, syntax correctness, semantic alignment, execution accuracy, and robustness, while also recording resource consumption and error distribution. Utilizing this framework, this study conducts a systematic evaluation of 24 representative LLMs, covering six categories: general non-reasoning models, general reasoning-enhanced models, general code generation models, geospatial code generation models, general SQL generation models, and GeoSQL generation strategies.

### 5.1. Core conclusions and insights

The comprehensive experimental results indicate that different large models exhibit multidimensional differences in performance for GeoSQL code generation tasks. These differences can be summarized across five aspects: task structure, model type, capability level, error types, and overall efficiency.

- **Task Dimension**: In different task types, multiple-choice and true/false questions generally achieve higher accuracy, reflecting a mature conceptual understanding of the models. However, significant differences are observed in syntax-level SQL generation question and table schema retrieval question tasks: the former frequently makes errors in function calls and parameter matching, while the latter, though more stable in field recognition, experiences a substantial increase in syntax error rate due to complex logic and multiple table joins. Execution pass rate and overall accuracy emerge as key indicators for distinguishing model performance, with table schema retrieval remaining the primary bottleneck for current models.

- **Model Type**: General non-reasoning models and general reasoning-enhanced models perform exceptionally well in most tasks, with GPT-5, o4-mini, and DeepSeek-R1-0528 ranking at the top, balancing both accuracy and stability. Reasoning-enhanced models show clear advantages in complex tasks, although there is significant variation between models; for instance, GPT-OSS-20B and QwQ-32B exhibit suboptimal efficiency and accuracy. General SQL generation models demonstrate a scale effect, with larger models (e.g., XiYan-SQL-32B) performing significantly better than smaller models (e.g., CodeS series).

- **Capability Level**: At the conceptual understanding layer, general rule compliance accuracy is generally low, indicating insufficient understanding of the intricacies of PostGIS. At the structured SQL generation layer, syntax correctness does not guarantee high executability, as it is still constrained by the specific logic of PostGIS. The semantic alignment and invocation layer performs poorly, with low function hit rates and parameter matching accuracy, becoming the primary bottleneck. In terms of execution and answer accuracy, geometric tasks stand out, while numerical and textual tasks have notable shortcomings. Generalization and robust reasoning layer shows that general reasoning-enhanced models have an advantage in complex tasks, but their performance significantly deteriorates under conditions of semantic ambiguity, highlighting insufficient robustness and semantic disambiguation.

- **Error Types**: The errors are mainly concentrated in PostGIS function errors and SQL syntax errors, with each accounting for approximately 70%. There are also some errors in geometric analysis and SRID/dimension mismatches, with the accuracy of PostGIS function knowledge and syntax remaining a key focus for optimization.

- **Overall Efficiency**: In terms of execution efficiency, general non-reasoning models perform the best. General reasoning-enhanced models show significant differences, with o4-mini striking the best balance between accuracy and efficiency, achieving optimal performance. GPT-5 adapts reasoning length based on task difficulty, balancing both accuracy and resource consumption. Gemini 2.5 Flash-0520, as a hybrid reasoning model, also performs well. The cost-performance ratio becomes a key factor in determining the practical application value of the models.

**5.2. Significance and Advantages**

The GeoSQL-Eval evaluation framework proposed in this study extends the existing evaluation system for LLMs in the field of geospatial SQL query generation across multiple dimensions, including task chain integrity, capability level progression, and the granularity of evaluation metrics. This framework comprehensively covers the entire task process from language understanding and syntax generation to pattern matching and execution validation, enhancing the accuracy in characterizing the model's transferability, generalization ability, and robustness in GeoSQL scenarios. Compared to traditional evaluation methods, which often suffer from limited task types, small sample sizes, and reliance on manual judgment, GeoSQL-Eval utilizes an end-to-end automated evaluation mechanism and a multidimensional metric system. It can capture semantic bias and code hallucinations, and provide quantifiable and interpretable analysis of model behavior.

Practically, the GeoSQL-Bench benchmark set constructed in this study covers three major task categories: multiple-choice and true/false questions, syntax generation questions, and table schema retrieval questions. It balances function identification, syntax structure, and table schema invocation, comprehensively reflecting the high complexity, strong professionalism, and diversity inherent in geospatial queries. This not only provides an effective path for practical screening and deployment risk control of models but also lays the foundation for the optimization and implementation of future GeoAI applications. On the technical level, GeoSQL-Eval achieves the stability, controllability, and reproducibility of the evaluation process based on the local operating mechanism of the PostGIS database engine, with low costs for data and framework migration. For model development, GeoSQL-Eval provides fine-grained feedback for both general models and specialized GeoSQL models, supporting model training and fine-tuning, and driving the evolution of LLMs into geospatial intelligent agents. GeoSQL-Eval not only extends the functional boundaries of existing tools but also provides a systematic measurement method for the credibility, professional adaptability, and cross-scenario application potential of LLMs, offering significant academic value and engineering application significance.

### 5.3. Limitations and Future Work

Although the GeoSQL-Eval framework proposed in this study achieves systematic expansion in task coverage, hierarchical design, and the metric system, there are still areas that need improvement. Firstly, this study does not cover all the functions listed in the official PostGIS documentation. The current test set mainly focuses on vector processing functions, such as Spatial Functions and Operators, while other functions like raster processing, network analysis, and advanced spatial operations modules have not yet been included. In the future, the function scope could be further expanded to enhance the dataset's completeness and representativeness. Secondly, the evaluation in this study focuses on the PostGIS plugin itself. However, in practical applications, there are numerous PostGIS-based extension plugins and other spatial database systems, which could be gradually incorporated into future research to expand the framework's applicability and generalizability. In terms of model selection, this study evaluates the performance of a representative set of the latest LLMs, which reflects the overall level of current mainstream models. However, there is still a problem of insufficient coverage. Future work could introduce more types and scales of models to enable a broader and deeper performance comparison. Additionally, future research could explore a unified evaluation mechanism across tasks, modalities, and platforms, and integrate with dynamically updated open platforms. This would allow research teams to upload models for continuous evaluation and iterative optimization, thereby establishing a long-term sustainable geospatial intelligence evaluation infrastructure.

**ORCID**

Shuyang Hou: https://orcid.org/0009-0000-6984-9959
Haoyue Jiao: https://orcid.org/0009-0000-5699-3919
Ziqi Liu: https://orcid.org/0009-0001-0483-5502
Lutong Xie: https://orcid.org/0009-0007-3927-7129
Guanyu Chen: https://orcid.org/0009-0002-9016-9143
Shaowen Wu: https://orcid.org/0009-0000-6875-9900
Xuefeng Guan: https://orcid.org/0000-0003-0865-3850
Huayi Wu: https://orcid.org/0000-0003-3971-0512

**Funding**

The work was supported by the National Natural Science Foundation of China [Grant number 41930107].

**CRediT authorship contribution statement**

**Shuyang Hou**: Writing – review & editing, Writing – original draft, Visualization, Methodology, Formal analysis, Conceptualization. **Haoyue Jiao**: Writing – review & editing, Writing – original draft, Methodology, Formal analysis, Data curation, Visualization, Software. **Ziqi Liu**: Validation, Software, Project administration, Investigation, Data curation. **Lutong Xie**: Software, Resources, Formal analysis, Data curation. **Guanyu Chen**: Investigation, Data curation, Validation. **Shaowen Wu**: Investigation, Data curation. **Xuefeng Guan**: Validation, Resources, Supervision. **Huayi Wu**: Supervision, Funding acquisition, Conceptualization.

**Declaration of competing interest**

The authors declare the following financial interests/personal relationships which may be considered as potential competing interests: Huayi Wu reports financial support was provided by National Natural Science Foundation of China (No.41930107). If there are other authors, they declare that they have no known competing financial interests or personal relationships that could have appeared to influence the work reported in this paper.

**Acknowledgements**

This work was supported by the National Natural Science Foundation of China under Grant No. 41930107, awarded to Huayi Wu.

**Appendix A Thematic Database Overview**

Appendix A corresponds to **Section 2.2.3 *Table Schema Retrieval Question*** and provides a detailed overview of the constructed thematic databases. Specifically, Table A1 presents the comprehensive composition of each database, including the thematic category, the number of tables contained, and the distribution of fields within these tables.

**Table A1 Comprehensive Composition of Thematic Databases and Statistics of Tables and Fields.** Topic Category IDs correspond to those in Table 2, Section 2.2.3

| Database Name | Table Count | Field Count | Topic Category ID | Database Name | Table Count | Field Count | Topic Category ID |
|---|---|---|---|---|---|---|---|
| AddressFabric | 1 | 6 | 8 | LandUsePlanner | 5 | 35 | 3 |
| AeroFlowCity | 7 | 53 | 2 | LithoBase | 1 | 6 | 2 |
| AgriPollutionTrace | 8 | 36 | 6 | LuminaGrid | 3 | 33 | 4 |
| AgroForestBase | 6 | 36 | 1 | MarineActivityMonitor | 3 | 15 | 2 |
| AquaFlowNet | 6 | 43 | 2 | MarineHabitatMapper | 4 | 20 | 2 |
| AquaLens | 3 | 15 | 2 | MarshEcoWatch | 6 | 48 | 1 |
| AquaTrendWatch | 8 | 35 | 2 | MicroClimateZones | 4 | 19 | 2 |
| BioDiversityAtlas | 5 | 41 | 1 | MineRehabAssessment | 7 | 49 | 3 |
| BuildingFootprintRegistry | 1 | 6 | 4 | NoctaLume | 7 | 49 | 5 |
| CarePathTracker | 8 | 54 | 5 | OrbitalEncroachmentDetector | 4 | 18 | 7 |
| CityFringeWatch | 7 | 49 | 3 | OrbitalRoadNet | 1 | 6 | 7 |
| CityReflex | 6 | 37 | 4 | ParkingPulse | 7 | 49 | 5 |
| CityScapeAdmin | 6 | 36 | 4 | PathogenTrail | 4 | 18 | 7 |
| CrimeHotspotTracker | 8 | 52 | 5 | PathogenTrajectory | 3 | 30 | 6 |
| DensityPulse | 7 | 49 | 5 | PatrolNet | 6 | 38 | 4 |
| EcoEnforceHub | 8 | 55 | 6 | PatrolResponseNet | 5 | 31 | 4 |
| EcoGuardian | 6 | 36 | 6 | PedosphereQualityGrid | 4 | 16 | 1 |
| EcoVigilance | 5 | 35 | 6 | PixelShiftArchive | 1 | 6 | 7 |
| EduSpatialBalance | 8 | 36 | 3 | PowerGridVigil | 6 | 36 | 4 |
| EstatePulse | 6 | 47 | 3 | PropertyValueCore | 7 | 42 | 3 |
| FarmlandGuardian | 6 | 47 | 3 | RedZonePerimeter | 2 | 10 | 4 |
| FiberTrenchVision | 2 | 11 | 4 | RuralHomesteadRegistry | 7 | 49 | 8 |
| FireCoverageOptimizer | 8 | 44 | 6 | SafeHavenSiteOptimizer | 8 | 49 | 6 |
| ForestCanopy | 6 | 42 | 1 | ServiceFootprint | 6 | 36 | 5 |
| ForestCanopyWatch | 6 | 35 | 1 | ServiceFootprint-2 | 5 | 35 | 5 |
| ForestDynamicsDB | 5 | 35 | 1 | SilverCareCommunity | 8 | 46 | 5 |
| GrassVitality | 6 | 40 | 1 | SlopeHazardMapper | 4 | 17 | 1 |
| GreenCanopyMaster | 7 | 49 | 4 | SlopeHazardSampler | 1 | 5 | 3 |
| HavenPathDB | 3 | 31 | 6 | SoundScapeRegulator | 7 | 50 | 6 |
| HazardResponseCore | 5 | 35 | 6 | StormReadyCity | 4 | 40 | 6 |

| Database Name | Table Count | Field Count | Topic Category ID | Database Name | Table Count | Field Count | Topic Category ID |
|---|---|---|---|---|---|---|---|
| HazardResponseNet | 6 | 41 | 6 | SubterrainMaster | 7 | 49 | 4 |
| HealthAccessNet | 8 | 35 | 5 | TerrainProfile | 1 | 5 | 7 |
| HealthCoverageGrid | 2 | 8 | 4 | ThermalScape | 7 | 49 | 2 |
| HeritageFabric | 7 | 47 | 8 | ThermoHumidityAlert | 3 | 14 | 2 |
| HeritageGuard | 3 | 40 | 8 | TideWatch | 4 | 16 | 2 |
| HeritagePathfinder | 5 | 33 | 8 | ToponymEvolution | 1 | 6 | 8 |
| HeritageRevitalization | 6 | 38 | 3 | ToponymRegistry | 1 | 6 | 8 |
| InfraVigil | 5 | 37 | 4 | TrafficFlowInsight | 6 | 40 | 5 |
| LandRejuvenationHub | 7 | 49 | 3 | TrafficPulse3D | 5 | 30 | 5 |
| LandUseIntensityHub | 7 | 41 | 3 | TrailHealthNetwork | 3 | 29 | 3 |
| LandUseMaster | 6 | 37 | 3 | WildSanctuaryGuard | 6 | 52 | 1 |